\documentclass[12pt]{article}

\usepackage{axodraw,cite}

\newcommand{\p}{\psi}

\newcommand{\be}{\begin{eqnarray}}   
\newcommand{\en}{\end{eqnarray}}

\newcommand{\csch}{\mathrm{\;csch\;}}

\begin{document}

\begin{center}

\begin{small}
\hfill FTUAM-05-3 \\
\hfill IFT-UAM/CSIC-05-15 \\
\hfill NEIP-05/03\\
\hfill ROMA-1385/04 \\
\hfill TUM-HEP-578/05 \\
\end{small}

\vspace{0.8 cm}

{\LARGE\bf Minimal gauge-Higgs unification\\[2mm]
with a flavour symmetry}

\vspace{1.2cm}

{\bf G. Martinelli$^{a}$, M. Salvatori$^{b}$, C. A. Scrucca$^{c}$,
L. Silvestrini$^{a,d}$}\\

\vspace{1.2cm}

${}^a\!\!$
{\em Dip. di Fisica, Univ. di Roma ``La Sapienza'',  and INFN, Sez. di Roma, } \\
{\em Piazzale Aldo Moro 2, 00185 Rome, Italy}
\vspace{.3cm}

${}^b\!\!$
{\em Dep. de F\'\i sica Te\'orica, C-XI, and 
Ist. de  F\'\i sica Te\'orica, C-XVI} \\ 
{\em Univ. Aut\'onoma de Madrid, Cantoblanco, 28049 Madrid, Spain}
\vspace{.3cm}

${}^c\!\!$
{\em Institut de Physique, Univ. de Neuch{\^a}tel, Rue Breguet 1,\\ 
2000 Neuch{\^a}tel, Switzerland}
\vspace{.3cm}

${}^d\!\!$
{\em Physik-Department T31, TU-M{\"u}nchen, 85748 Garching, Germany}
\vspace{.3cm}

\end{center}

\vspace{0.8cm}

\centerline{\bf Abstract}
\vspace{2 mm}
\begin{quote}\small

  We show that a flavour symmetry {\`a} la Froggatt-Nielsen can be
  naturally incorporated in models with gauge-Higgs unification, by
  exploiting the heavy fermions that are anyhow needed to realize
  realistic Yukawa couplings.  The case of the minimal
  five-dimensional model, in which the $SU(2)_L \times U(1)_Y$
  electroweak group is enlarged to an $SU(3)_W$ group, and then broken
  to $U(1)_{\rm em}$ by the combination of an orbifold projection and
  a Scherk-Schwarz twist, is studied in detail. We show that the
  minimal way of incorporating a $U(1)_F$ flavour symmetry is to
  enlarge it to an $SU(2)_F$ group, which is then completely broken by
  the same orbifold projection and Scherk-Schwarz twist. The general
  features of this construction, where ordinary fermions live on the
  branes defined by the orbifold fixed-points and messenger fermions
  live in the bulk, are compared to those of ordinary four-dimensional
  flavour models, and some explicit examples are constructed.

\end{quote}

\newpage

\section{Introduction}

In the last thirty years the central problem in particle physics
has been the mechanism for breaking the electroweak gauge
symmetry and the consequent generation of masses for gauge bosons
and matter fermions. In the Standard Model (SM), the problem manifests
itself in two different ways: on the one hand in the instability of
the weak interaction scale (the so-called hierarchy problem), on the
other in the arbitrariness of the Yukawa couplings, which span at least
five orders of magnitude, and the related problem of the strength of
the CKM~\cite{ckm} (and PMNS~\cite{pmns}) matrix elements. 
In the quest for a solution of these fundamental problems, a plethora 
of extensions of the SM have been proposed, like technicolor, softly broken 
global supersymmetry, supergravity or string theory. None of the proposed
solutions, however,  is satisfactory, and this motivates further 
investigation.

More recently the idea came on the stage that it might be possible to
overcome the hierarchy problem by implementing the breaking of gauge 
symmetries via alternative mechanisms relying on the presence of one 
or more extra dimensions. In particular, it has been known for a
long time that realizing the Higgs field as the zero-mode of the 
internal component of a higher-dimensional gauge field leads to 
an effective potential with improved stability \cite{gaugehiggsgeneral}.
This idea has been recently reinvestigated from various points of view
\cite{gaugehiggsrecent}, and exploited to construct concrete 
higher-dimensional orbifold models with this type of gauge-Higgs 
unification \cite{gaugehiggsorb,sss,sssw,gaugehiggsorb2,gaugehiggsorb3} 
(see also \cite{gaugehiggsorbsusy,burdman,gaugehiggsorbsusy2} 
for supersymmetric models). A simple prototype of this kind of models 
is the minimal five-dimensional (5D) scenario described in~\cite{sss}, 
where the gauge symmetry is broken by the combination of a $\mathbf{Z}_{2}$ 
orbifold projection \cite{orbifoldoriginale} and a continuous 
Scherk-Schwarz (SS) twist \cite{ss} along the extra compact dimension. 
The electroweak symmetry breaking is spontaneous and occurs through 
the Hosotani mechanism \cite{Hosotani}. The order parameter is the 
Wilson loop $W = \exp\,\{ i g \oint \! A_5(y) dy\}$ of the 
internal component $A_5$ of the gauge field along the internal circle 
$S^1$ (here and in the following we denote by $y$ the coordinate along the
internal dimension). The role of the Higgs field is played by the zero-mode of
$A_5$, but the effective potential can depend only on the non-local
gauge-invariant $W$ and is therefore finite.

In this paper, we study the possibility of endowing the
above-mentioned class of higher-dimensional models with a flavour
symmetry of the Froggatt-Nielsen (FN) type \cite{froggatt}. This is
done by introducing an extended flavour symmetry, which is then
broken, as for the electroweak symmetry, by the combination of an
orbifold projection and a SS twist.  We focus on the model of
ref.~\cite{sss} and describe its minimal flavour extension. We show
that by a wise choice of the flavour quantum numbers for bulk and
brane fermion fields, it is possible to reproduce the observed pattern
of the quark masses and CKM angles, although the mass obtained for the
down quark tends to be too small, and observe that a similar approach
is possible also for lepton masses and PMNS angles. The resulting
model generates a 4D effective theory with a stabilized electroweak
scale and a $U(1)$ FN symmetry.

The paper is organized as follows. After quickly reviewing gauge-Higgs
unification in sec.~\ref{ghu}, we outline in sec.~\ref{sec:prototype} 
the basic construction discussing explicitly a prototype model. In
sec.~\ref{sec:perglistringhisti} we generalize our construction to arbitrary
representations of the electroweak and flavour groups. In sec.~\ref{minimal} 
we present more realistic examples of our construction. Finally, in 
sec.~\ref{conclusion} we draw some conclusions and discuss 
future developments.

\section{Gauge-Higgs unification in 5D}
\label{ghu}

Our starting point is the model of gauge-Higgs unification described in
ref.~\cite{sss}. The basic physical idea is to break the electroweak symmetry
in a non-local way, so that the Higgs mass is protected by the gauge invariance 
of the 5D theory. Indeed, in a 5D theory compactified on a circle with SS symmetry
breaking, all ultraviolet (UV) divergent quantities at all orders in perturbation 
theory must be invariant under the full symmetries of the 5D theory \cite{msss}. 
This means that all symmetry-breaking quantities are finite, calculable
and insensitive to the unknown UV dynamics. If one could find a 5D symmetry that
forbid the Higgs mass term, a non-local breaking of this symmetry would protect
the Higgs mass from any divergent radiative correction. Gauge-Higgs unification
implements this idea, by identifying the Higgs boson with the internal
component of a 5D gauge field, so that the 5D gauge symmetry protects the Higgs
mass.
 
To construct a model of gauge-Higgs unification one must consider a
gauge group large enough as to include 4D states corresponding to the $SU(2)_L
\otimes U(1)_Y$ gauge bosons plus the Higgs doublet.  The minimal possibility
corresponds to an $SU(3)_W$ gauge group, broken first to $SU(2)_L \otimes U(1)_Y$
via a $Z_2$ orbifold projection, and then to $U(1)_\mathrm{em}$ with a SS
twist.\footnote{In this way, one obtains 
$\sin^2 \theta_W = 3/4$. An acceptable value of the weak mixing angle 
can be achieved by adding an extra $U(1)^\prime$ factor and tuning its
coupling relatively to the weak coupling, as done in ref.~\cite{sss}. 
The additional $U(1)_X$ symmetry introduced in this way in the 4D effective 
theory is anomalous, and must therefore be spontaneously broken and decoupled.}
The orbifold projection acting on the 5D gauge group leaves as 4D zero modes the
SM gauge bosons plus a scalar doublet with the quantum numbers of the SM Higgs:
the SS twist corresponds to a Vacuum Expectation Value (VEV) for the Higgs via the 
Hosotani mechanism.
From the 4D point of view, this corresponds to the SM Higgs mechanism: however,
higher-dimensional gauge invariance protects the Higgs mass. This remains true
even though at the orbifold fixed points only the SM gauge group is present:
indeed,  the zero-modes of $A_5$ 
transform non-homogeneously under gauge transformations belonging to
$SU(3)_W/(SU(2)_L\otimes U(1)_Y)$, so that the only possible counterterms are
$SU(3)_W$-invariant ones~\cite{quiros} (see also \cite{nonpert}).

The price one has to pay for this UV insensitivity is the absence of a
tree-level potential for the Higgs. This implies that the Higgs mass generated
at one loop is generically too small (see ref.~\cite{sss} for a detailed
discussion of this problem). A related issue is the value of the SS twist that is
dynamically generated: unless bulk fermions belonging to very high-rank
representations of $SU(3)$ are present, one obtains twist parameters of order $10^{-1}$,
corresponding to an extra dimension of inverse radius $1/R \sim 10\, m_W \sim 1$
TeV, far below the LEP indirect bounds. Since these problems are unrelated to
the issue of flavour symmetry breaking that will be discussed in this work, from
now on we will assume that the value of the SS twist $\alpha$ is of order 
$10^{-2}$ thanks to some unspecified mechanism, so that $1/R \sim 10$ TeV 
and Kaluza-Klein (KK) excitations of electroweak gauge bosons do not pose 
any phenomenological problem. 

As in the standard electroweak theory, the VEV of the Higgs field can 
induce a mass for the matter fermions. The relevant Yukawa couplings,
however, originate in this case from the 5D gauge coupling. For bulk
fermions, this implies that the Yukawa couplings are universal and their 
magnitude is simply the gauge coupling times a group-theoretical factor, 
depending only on the representation. Furthermore, no flavour symmetry breaking
can arise from electroweak gauge couplings, so that one is left with a
universal fermion mass and no flavour mixing. 
For brane fields localized at the orbifold fixed-points, on the contrary, 
the $SU(2)_L \times U(1)_Y$ symmetry would allow Yukawa couplings to be 
arbitrary and non-universal, but the non-linearly realized 
$SU(3)_W/(SU(2)_L \times U(1)_Y)$ symmetry implies
that they all vanish. In order to achieve realistic Yukawa couplings, one is
therefore led to consider the more general case of fermions that are a mixture
of bulk and brane fields with wave functions depending non-trivially on the
internal dimension \cite{burdman}. This situation is most easily realized by
considering bulk and brane fields that mix through non-universal bilinear
couplings localized at the fixed-points \cite{sss}. The new eigenstates,
resulting from the diagonalization of the quadratic Lagrangian for
these fields, will then inherit non-vanishing and non-universal Yukawa
couplings to the Higgs field. The structure of the mass couplings is  pretty
general, but their size is always at most of the order of the gauge coupling. 
This implies that the natural value of all the fermion masses induced in
this way is of the order of $m_W$.

In the case where the above construction is realized with bulk fields 
that are much heavier than the brane fields, the lightest eigenstates 
are sharply localized fields whose dynamics is well approximated by an 
effective Lagrangian for the original brane fermions, obtained by 
integrating out the heavy bulk modes. From this perspective, the 
non-vanishing and non-universal Yukawa couplings for the light localized 
modes emerge as effective interactions induced through the exchange of 
the heavy bulk fermions, which have a non-vanishing but universal 
fundamental Yukawa coupling. This framework is very similar to the one 
occurring in models with flavour symmetries, the breaking of which is 
transmitted to the effective Yukawa couplings through a heavy fermion,
and suggests that it should be possible to naturally generalize the model 
of ref.~\cite{sss} to include a flavour symmetry. 

The usual implementation of a FN $U(1)_F$ flavour symmetry goes as follows. One
assigns a $U(1)_F$ charge to each of the SM fermions, and introduces some heavy 
vector-like fermions in order to construct gauge- and flavour-invariant Yukawa couplings.
The flavour symmetry is then spontaneously broken by some VEV at a scale smaller
than the mass of the heavy fermions, so that the effective Yukawa couplings for
SM fermions generated at low energies are $Y_{IJ} \propto ( \langle \phi
\rangle /M )^{q_I - q_J}$, where $\langle \phi \rangle$ is the  $U(1)_F$-breaking
VEV, $M$ is the mass of the heavy fermions, and $q_{I}$ are the SM fermion flavour
charges. Wave-function corrections and the potentially dangerous tree-level FCNC
generated by the heavy fermions are power suppressed and negligible if the new
particles live at a high scale. 

It is then natural for us to consider the case of a $U(1)_F$ symmetry broken \`a
la SS. Since rank lowering can only be achieved by combining an orbifold
projection with a SS twist, we have to start from an $SU(2)_F$ symmetry in the
bulk, broken to $U(1)_F$ by the orbifold and then to nothing via a SS twist.
Clearly, since the mass scale of the heavy bulk fermions is around $10$ TeV in
the case of ref.~\cite{sss}, we should make sure that wave function corrections
and tree-level FCNC couplings are under control.  We have performed a
preliminary analysis of this issue, which indicates that unwanted effects might
indeed be kept sufficiently small with reasonable choices of parameters. A detailed
analysis, together with a study of loop-induced FCNC's, is currently under way
and will be presented elsewhere.

\section{A prototype model}
\label{sec:prototype}

A minimal prototype of the models discussed in the previous section can be
constructed as follows.
The standard fermions are taken to live at the orbifold fixed-points, 
whereas the messenger fermions that activate the mechanism of symmetry 
breaking live in the bulk. A spontaneously broken Abelian flavour 
symmetry is then incorporated much in the same way as for the spontaneously 
broken electroweak symmetry, and both symmetry breakings are implemented 
at once by letting the orbifold projection and the SS twist act on 
both the electroweak and the flavour groups. The minimal choice of 5D
flavour group allowing an Abelian group in the intermediate step and a full 
breaking in the final step is an $SU(2)_{F}$ group. For simplicity, we 
assume this to be a global symmetry, but the case of a local symmetry is 
similar. This flavour group is broken to a $U(1)_F$ subgroup through the
orbifold projection, and finally to nothing through the SS twist. 

The above construction is very general, and exploits for both the 
electroweak and the flavour symmetries the same minimal pattern of 
symmetry breaking discussed in ref.~\cite{bhn}, which consists in first 
promoting the 4D group to a larger 5D group and then performing two 
non-commuting projections that enable to lower the rank. The standard 
fermions at the fixed-points form representations of 
$SU(2)_L \times U(1)_Y \times U(1)_F$, whereas the messenger fermions 
in the bulk form representations of $SU(3)_W \times SU(2)_F$. 
The construction can be applied in a perfectly similar way both to the 
quark and the lepton sectors. Here we shall focus on the quark sector. 
For the sake of clarity of presentation, we will first illustrate the general 
qualitative features of the construction with an explicit example, then
generalize to arbitrary flavour charges and $SU(3)_W \otimes SU(2)_F$ 
representations, and finally discuss some realistic models.

\subsection{Orbifold projection and SS twist}

The projections defining the model are chosen as follows. The orbifold 
projection on a bulk field $\Phi_{{\cal R},{\cal R}^\prime}$ in a generic 
representation $({\cal R},{\cal R}^\prime)$ of $SU(3)_W \times SU(2)_F$ 
is taken to be
\begin{eqnarray}
\Phi_{{\cal R},{\cal R}^\prime}(x,-y) = \pm \big[P_L \otimes P_W^{\cal R}
\otimes 
P_F^{{\cal R}^\prime}\big]\, \Phi_{{\cal R},{\cal R}^\prime}(x,y) \,,
\label{eq:orbi}
\end{eqnarray} 
where $P_L$ depends on which Lorentz representation the field corresponds
to ($P_L=1$ for a scalar, $P_L = \gamma_5$ for a spinor, etc.) and $P_W$ 
and $P_F$ define the embedding of the projection into the weak and flavour 
groups. In order to achieve the desired symmetry breaking down to 
$SU(2)_W \times U(1)_Y \times U(1)_F$, we use the $T_W^8$ and $T_F^3$ 
generators\footnote{We define the $SU(3)_{W}$ generators as 
$T^{a} = \lambda^{a}/2$, where $\lambda^{a}$ are the standard 
Gell-Mann matrices with the normalization $\mathrm{Tr} \lambda^{a}
\lambda^{b} = 2 \delta^{ab}$. Similarly, we define the $SU(2)_{F}$ 
as $T^{a} = \sigma^{a}/2$, where $\sigma^{a}$ are the standard Pauli matrices
with the normalization $\mathrm{Tr} \sigma^{a} \sigma^{b} = 
2 \delta^{ab}$.} of $SU(3)_W$ and $SU(2)_F$ respectively, and choose:
\begin{eqnarray}
P_W = e^{2 i \pi \sqrt{3} T_W^8} \,,\qquad
P_F = e^{-i \pi\, (d(T_F^3)-1)/2}\, e^{i \pi T_F^3} \,,
\label{eq:orbidef}
\end{eqnarray}
where $d(T)$ is the dimension of the matrix $T$ acting on the representation
${\cal R}^\prime$.
The residual $SU(2)_F \times U(1)_Y$ electroweak gauge symmetries are 
associated with the generators $T^{a}_{W}$ with $a=1,2,3,8$ that 
commute with the projection: $[T^{a}_{W},P_W] = 0$.
Similarly, the surviving $U(1)_F$ flavour symmetry is associated 
to the only generator $T^{3}_{F}$ commuting with the projection: 
$[T^{3}_{F},P_F] = 0$. 

The Scherk-Schwarz twist on the generic representation $({\cal R},{\cal R}^\prime)$ 
of $SU(3)_W \times SU(2)_F$ is similarly of the form:
\begin{eqnarray}
\Phi_{{\cal R},{\cal R}^\prime}(x,y+ 2 \pi R) = \left[ T_W^{\cal R}(\alpha)
\otimes 
T_F^{{\cal R}^\prime}(\beta)\right] \, \Phi_{{\cal R},{\cal R}^\prime}(x,y) \,, 
\label{eq:twist}
\end{eqnarray} 
where $T_W(\alpha)$ and $T_F(\beta)$ define the embedding of the twist 
into the weak and flavour groups and depend on two continuous parameters 
$\alpha$ and $\beta$. These must satisfy the usual consistency constraints
$(T_W P_W)^2 = (T_F P_F)^2 = 1$ \cite{orbifoldoriginale,relazioneconsistenza}.
In order to further break by the twist the electroweak and flavour symmetries 
$SU(2)_F \times U(1)_Y \times U(1)_F$ preserved by the orbifold projection down 
to $U(1)_{\rm em}$, we use the $T_W^6$ and $T_F^1$ generators of $SU(3)_W$ and 
$SU(2)_F$ respectively, and choose:
\begin{eqnarray}
T_W(\alpha) = e^{4 \pi i \alpha T_W^6}\,,\qquad
T_F(\beta) = e^{4 \pi i \beta T_F^1} \,.
\label{eq:twistdef}
\end{eqnarray}
The residual $U(1)_{\rm em}$ electromagnetic gauge symmetry is 
associated with the only generator $T^{3}_{W}+T^8_W/\sqrt{3}$ that commutes 
also with the twist: $[T^{3}_{W}+T^8_W/\sqrt{3},T_W] = 0$. Notice that this
fixes the hypercharge to be $Y=T^8_W/\sqrt{3}$. The flavour symmetry 
is instead completely broken since there is no generator commuting
also with the twist.

The dimensionless quantities $\alpha$ and $\beta$ are the order parameters
for the rank-reducing breaking of the electroweak and flavour symmetries.
Indeed, it is evident from eqs.~(\ref{eq:orbidef}) and (\ref{eq:twistdef})
that the orbifold projection and the twist do not commute, that is 
$[P_W,T_W] \neq 0$ in the gauge sector and $[P_F,T_F] \neq 0$ in the flavour 
sector, unless $\alpha = n/2$ and $\beta = n/2$, with $n$ integer. 
For the gauge symmetry, it is possible to relate the order parameter 
to the VEV of the Higgs field $A_5$ by performing a non-periodic 
gauge transformation that reabsorbs the twist \cite{Hosotani}:
$\alpha = g_5 R \langle A_5 \rangle/2$. For the flavour symmetry, a similar 
relation would hold if it were local; the case where it is taken to
be global can however be understood in a similar way by taking a 
suitable decoupling limit \cite{msss}. Notice finally that the 
electroweak and flavour symmetry breaking scales are naturally 
defined by $m_W = \alpha/R$ and $m_F = \beta/R$.

The effect of the SS twist on the orbifold-projected spectrum of 
KK modes of bulk fields will as usual amount to 
shifting the standard integer-moded masses $m_n = n/R$ obtained for 
fields that are periodic along the internal circle $S^1$ with 
radius $R$ through a quantity that depends on the symmetry 
breaking parameters $\alpha$ and $\beta$. To be more precise,
notice that the generators appearing in the exponents of the 
orbifold projection and SS twist do not commute. Starting from
the standard basis in which the Cartan generators $T_W^8$ and $T_F^3$ 
appearing in the orbifold projection are diagonal, the generators
$T_W^6$ and $T_F^1$ appearing in the twist can be brought into 
diagonal forms, which we denote by $t_W$ and $t_F$, through some 
suitable unitary transformations $U_W$ and $V_F$:
\begin{eqnarray}
t_W = U_W T_W^6 U_W^\dagger \,,\qquad
t_F = V_F T_F^1 V_F^\dagger \,.
\label{diag}
\end{eqnarray}
In the transformed basis where the SS twist is diagonal (but the 
orbifold projection is not diagonal), the mass spectrum can be written 
in terms of the entries of the diagonalized twist generator simply as 
$m_n(\alpha,\beta) = (n + 2 t_W \alpha + 2 t_F \beta)/R$ (see
sec.~\ref{modedecom}).

\subsection{Field content}

The field content of the model is a generalization of the one 
considered in ref.~\cite{sss}, where now all the brane fields 
must not only belong to $SU(2)_L \times U(1)_Y$ representations but also 
have definite charges under the $U(1)_F$ subgroup, and similarly 
all the bulk fields must also belong not only to $SU(3)_W$ but also to
$SU(2)_F$ representations. Notice that the charge under the $U(1)_F$ flavour 
group preserved by the orbifold projection is quantized, as a 
consequence of the fact that the original flavour group is non-Abelian, 
and represented by $q_{F} = T^{3}_F$. This constrains in an interesting
way the allowed charge assignments for the brane fields.
The minimal content of brane and bulk fields that is required in 
order to construct the flavour extension of the model of ref.~\cite{sss} 
is then quite rigidly fixed. 

The SM fermions are introduced as brane fields at the fixed-points of 
the orbifold projection. Denoting by $y$ the periodic coordinate of 
the extra dimension, the two fixed-points are located at $y=0$ and 
$y=\pi R$ and represent the two boundaries of the physical space, 
the segment $[0,\pi R]$ in the extra dimension. 
Each of the left- and right-handed fields can be located at any of the 
two fixed-points. The precise distribution that is chosen is qualitatively 
not too important as far as the low-energy effective theory is concerned, 
but it is quite relevant for the consistency of the theory, and in particular 
for the issue of anomalies. Indeed, it is known that globally vanishing 
localized anomalies occur in theories with a generic content of bulk and 
brane fields and that requiring their cancellation may have non-trivial 
implications on the theory \cite{anoorb,localsymmetry} (see \cite{reviewano} 
for a general review). The issue of localized anomalies has already been 
discussed in ref.~\cite{sss}, and since the flavor extension examined here 
does not involve any novelty in this respect, we shall not discuss it any 
further here.
For simplicity, we assume that all the left-handed fields are located 
at $y=0$ and the right-handed ones at $y=\pi R$, and their interactions 
are constrained to be invariant under the residual symmetries described 
above. We introduce the following representations of $SU(2)_L \times U(1)_Y \times
U(1)_F$:

\begin{itemize} 
\item Left-handed fields localized at $y=0$:
\begin{eqnarray}
\begin{array}{ccc}
\vspace{1em}
Q_{L} = \left(
\begin{array}{c}
u_{L} \\
d_{L} 
\end{array}
\right) : {\bf 2}_{\frac{1}{6},q}\,\quad \mbox{or equivalently}\quad 
Q_{R}^{c} = \left(
\begin{array}{c}
d_{R}^{c} \\
\!\!\mbox{-}u_{R}^{c}
\end{array}
\right) =  {\bf 2}_{-\frac{1}{6},-q} \;.
\end{array}
\end{eqnarray}
\item Right-handed fields localized at $y = \pi R$: 
\begin{eqnarray}
\begin{array}{lll}
\vspace{0.5em}
u_{R} = {\bf 1}_{\frac{2}{3},u}\,,&  \mbox{or equivalently} & 
\!\!\mbox{-}u_{L}^{c} = {\bf 1}_{-\frac{2}{3},-u} \,, \\
d_{R} = {\bf 1}_{- \frac{1}{3},d}\,,&  \mbox{or equivalently} &
d_{L}^{c} = {\bf 1}_{ \frac{1}{3},-d}\,,
\end{array} 
\end{eqnarray}
\end{itemize}
with the notation $\mathbf{R}_{q_Y,q_F}$, where $\mathbf{R}$ is the $SU(2)_L$
representation and $q_Y$ and $q_F$ are the $U(1)_Y$ and $U(1)_F$ charges
respectively. As a first example, we choose the charge assignment reported in
Table \ref{caricheMS}.

\vskip 10pt
\begin{table}[h]
\begin{center}
\begin{tabular}{||l|r||l|r||l|r||}
\hline
Field & $q_{F}$ & Field & $q_{F}$ & Field & $q_{F}$\\ 
\hline
\hline 
$Q_{1L}$ & $4$ & $d_{1R}$ & $-1$ & $u_{1R}$ & $-4$\\
\hline
$Q_{2L}$ & $3$ & $d_{2R}$ & $0$ & $u_{2R}$ & $-1$   \\
\hline
$Q_{3L}$ & $1$ & $d_{3R}$ & $1$ & $u_{3R}$ & $1$ \\
\hline
\end{tabular}
\end{center}
\caption{\em \small Flavour charges of SM fermions.}
\label{caricheMS}
\end{table}

The bulk fields consist of the 5D gauge fields and of the heavy fermions 
that are needed to induce the effective Yukawa couplings as in 
ref.~\cite{sss}. The r\^{o}le played by the gauge fields has been 
extensively explained in ref.~\cite{sss} and will not be discussed 
again here. The only novelty concerns the heavy messenger fermions, 
which will now carry flavour quantum numbers. We introduce two pairs 
$l=u,d$ of fermion fields ($\p^{l},\tilde{\p}^{l}$) with opposite 
orbifold parities, with a bulk mass term that makes all their modes 
heavy. Following ref.~\cite{sss},
we take these two pairs to be weak triplets to generate
masses for down-type quarks, and weak sixplets to generate
masses for up-type quarks. Concerning the representation under
$SU(2)_F$, from Table~\ref{caricheMS} we see that $Q_L$ and
$u^c_L$ have flavour charges with absolute value up to four: the minimal choice
is therefore a nineplet, which contains fields
with $U(1)_F$ charges from $-4$ to $4$. Summarizing, we have bulk fields in the
following representations of $SU(3)_W \times SU(2)_F$:

\begin{itemize} 
\item Bulk fields with negative overall parity:
\begin{eqnarray}
\begin{array}{cc}
\displaystyle{\psi^d : 
\Big({\bf 3, 9 }\Big)} \,,&
\displaystyle{\tilde\psi^u : 
\Big({\bf 6, 9 }\Big)} \,,
\end{array}
\end{eqnarray}
\item Bulk fields with positive overall parity: 
\begin{eqnarray}
\begin{array}{cc}
\displaystyle{\tilde \psi^d : 
\Big({\bf 3, 9 }\Big)} \,, &
\displaystyle{\psi^u : 
\Big({\bf 6, 9 }\Big)} \,.
\end{array} 
\end{eqnarray}
\end{itemize}

The decomposition of the above representations of the $SU(3)_W \times SU(2)_F$
group under its $SU(2)_Y \times U(1)_Y \times U(1)_F$
subgroup, which we will need to determine the coupling of the bulk fields 
to the brane fields, has the following form:
\begin{eqnarray}
\Big({\bf 3}, {\bf 9}\Big) \!\!&\rightarrow\!\!& {\bf 2}_{\frac{1}{6},q} \oplus
{\bf 1}_{-\frac{1}{3},d}\,,  \nonumber \\
\Big({\bf 6}, {\bf 9}\Big) \!\!&\rightarrow\!\!& {\bf 3}_{\frac{1}{3},Q} \oplus {\bf
2}_{-\frac{1}{6},-q} \oplus
{\bf 1}_{-\frac{2}{3},-u}\,,
\end{eqnarray}
with $Q,q,u,d$ ranging from $-4$ to $4$. The only components that have the right
quantum numbers to couple 
to the brane fermions are the $SU(2)_W$ doublets and singlets, with $U(1)_F$
charges matching the SM ones given in Table \ref{caricheMS}.

The action of the orbifold projection on the bulk fermion 
fields is given by
\begin{eqnarray}
P^{\bf{3}}_W \!\!\!&=\!\!\!& \mathrm{diag}( 
-1 , -1 , 1)\; , \quad 
P^{\bf{6}}_W = \mathrm{diag}(1,1,-1,1,-1,1)\,,\nonumber \\
P^{\bf{9}}_F  \!\!\!&=\!\!\!& \mathrm{diag}(1,-1,1,-1,1,-1,1,-1,1)\,.
\end{eqnarray}
This implies that after the projection the particle content is given by
an electroweak doublet and an electroweak singlet
with flavour charges ranging from $-4$ to $4$, belonging to $\psi^l$
if the flavour charge is even  and to $\tilde \psi^l$ if it is odd. 
In other words, one and only one of the two bulk fields $\psi^l$ and 
$\tilde \psi^l$ always has a component with the right quantum numbers 
to couple to the SM brane fermions.

The choice of the $SU(3)_W$ representation for the messenger 
fermions in the bulk influences only the overall magnitude 
of the induced Yukawa couplings, whereas the choice of the 
$SU(2)_F$ representation for these bulk fermions, together 
with the $U(1)_F$ charges for the matter brane fermions, 
determines the flavour structure. 

\subsection{Lagrangian}

The structure of the Lagrangian is the same as in ref.~\cite{sss}: in addition
to the kinetic terms for the bulk and brane fields, we introduce an 
arbitrary bilinear mixing between them. The couplings of the three generations 
of left- and right-handed brane fields $Q_{L}, u_{R}, d_{R}$ and 
their conjugates to the bulk fields $\psi^l$ or $\tilde \psi^l$ are parametrized
by couplings
$e_L^{l}$ and $e_R^{l}$ with mass-dimension $1/2$, in each sector $l=u,d$. 
Each brane field can couple either to $\psi^l$ or $\tilde \psi^l$, and 
has therefore only one relevant coupling. To write these couplings more 
explicitly, it is convenient to embed the brane fields into new fields 
$\chi^{u,d}_{L,R}$,  $\tilde\chi^{u,d}_{L,R}$ which have the same matrix 
structure as the representations of $SU(3)_W \times SU(2)_F$ to which the 
bulk fields belong, the extra entries being filled with zeroes, and then further 
combine left and right components into Dirac fields: 
$\chi^{u,d} = \chi^{u,d}_{L} +  \chi^{u,d}_{R}$ and  
$\tilde\chi^{u,d} = \tilde\chi^{u,d}_{L} + \tilde\chi^{u,d}_{R}$.
Correspondingly, it is 
convenient to embed the diagonal matrices of couplings $e_1^{l}$ and 
$e_2^{l}$ in family space into new diagonal matrices of couplings 
$\hat e_1^{l}$ and $\hat e_2^{l}$ in flavour space. In our example,
we have:
\begin{eqnarray}
\chi^{d} \!\!\!&=\!\!\!& \left( \begin{array}{c}
1 \\
0 \\
0 \\
\end{array} \right)_W \!\!\!\otimes \left( \begin{array}{c}
u_L^1 \\ 
0 \\
0 \\
0 \\
0 \\
0 \\
0 \\
0 \\
0 \\
\end{array} \right)_F \!\!\!+ \left( \begin{array}{c}
0 \\
1 \\
0 \\
\end{array} \right)_W \!\!\!\otimes \left( \begin{array}{c}
d_L^1 \\ 
0 \\
0 \\
0 \\
0 \\
0 \\
0 \\
0 \\
0 \\
\end{array} \right)_F \!\!\!+ \left( \begin{array}{c}
0 \\ 
0 \\
1 \\
\end{array} \right)_W \!\!\!\otimes \left( \begin{array}{c}
0 \\
0 \\
0 \\
0 \\
d_R^2 \\
0 \\
0 \\
0 \\
0 \\
\end{array} \right)_F \label{chid} \,, \\
& & \nonumber \\
\tilde\chi^{d} \!\!\!&=\!\!\!& \left( \begin{array}{c}
1 \\
0 \\
0 \\
\end{array} \right)_W \!\!\!\otimes \left( \begin{array}{c}
0 \\ 
u_L^2  \\
0 \\
u_L^3  \\
0 \\
0 \\
0 \\
0 \\
0 \\
\end{array} \right)_F \!\!\!+ \left( \begin{array}{c}
0 \\
1 \\
0 \\
\end{array} \right)_W \!\!\!\otimes \left( \begin{array}{c}
0 \\ 
d_L^2 \\
0 \\
d_L^3  \\
0 \\
0 \\ 
0 \\ 
0 \\
0 \\
\end{array} \right)_F \!\!\!+ \left( \begin{array}{c}
0 \\
0 \\
1 \\
\end{array} \right)_W \!\!\!\otimes \left( \begin{array}{c}
0 \\ 
0 \\
0 \\
d_R^3  \\
0 \\
d_R^1  \\
0 \\
0 \\
0 \\
\end{array} \right)_F \label{chidt} \,, \\
& & \nonumber \\
\chi^{u}\!\!\!&=\!\!\!& \left( \begin{array}{cc}
0 \\
0 \\
0 \\
0 \\
1 \\
0 \\
\end{array} \right)_W \!\!\!\otimes \left( \begin{array}{c}
0 \\
0 \\
0 \\
0 \\
0 \\
0 \\
0 \\ 
0 \\
\!\mbox{-}u_R^{c1}\!\! \\
\end{array} \right)_F \!\!\!+ \left( \begin{array}{c}
0 \\
0 \\
1 \\
0 \\
0 \\
0 \\
\end{array} \right)_W \!\!\!\otimes \left( \begin{array}{c}
0 \\
0 \\
0 \\
0 \\
0 \\
0 \\
0 \\ 
0 \\
\!d_R^{c1}  \\
\end{array} \right)_F \!\!\!+ \left( \begin{array}{c}
0 \\
0 \\
0 \\
0 \\
0 \\
1 \\
\end{array} \right)_W \!\!\!\otimes \left( \begin{array}{c}
\!u_L^{c1}  \\ 
0 \\
0 \\
0 \\
0 \\
0 \\
0 \\
0 \\
0 \\
\end{array} \right)_F \label{chiu} \,, \\
& & \nonumber \\
\tilde\chi^{u}\!\!\!&=\!\!\!& \left( \begin{array}{cc}
0 \\
0 \\
0 \\
0 \\
1 \\
0 \\
\end{array} \right)_W \!\!\!\otimes \left( \begin{array}{c}
0 \\
0 \\
0 \\ 
0 \\
0 \\
\!\mbox{-}u_R^{c3}\!\! \\
0 \\
\!\mbox{-}u_R^{c2}\!\! \\
0 \\
\end{array} \right)_F \!\!\!+ \left( \begin{array}{c}
0 \\
0 \\
1 \\
0 \\
0 \\
0 \\
\end{array} \right)_W \!\!\!\otimes \left( \begin{array}{c}
0 \\
0 \\
0 \\ 
0 \\
0 \\
\!d_R^{c3} \\
0 \\
\!d_R^{c2} \\
0 \\
\end{array} \right)_F \!\!\!+ \left( \begin{array}{c}
0 \\
0 \\
0 \\
0 \\
0 \\
1 \\
\end{array} \right)_W \!\!\!\otimes \left( \begin{array}{c}
0 \\ 
0 \\
0 \\
\!u_L^{c2} \\
0 \\
\!u_L^{c3} \\
0 \\
0 \\
0 \\
\end{array} \right)_F\,, \label{chiut} 
\end{eqnarray}
with $u^{1,2,3}$ and $d^{1,2,3}$ denoting the three generation
quarks, and
\begin{eqnarray}
\begin{array}{lll}
\hat e^{d}_1 \!\!\!&=\!\!\!&
\mathrm{diag}(e^d_{1,1},e^d_{1,2},0,e^d_{1,3},0,0,0,0,0)\,,\\[1mm]
\hat e^{d}_2 \!\!\!&=\!\!\!&
\mathrm{diag}(0,0,0,e^d_{2,3},e^d_{2,2},e^d_{2,1},0,0,0)\,,\\[1mm]
\hat e^{u}_1 \!\!\!&=\!\!\!&
\mathrm{diag}(0,0,0,0,0,e^u_{1,3},0,e^u_{1,2},e^u_{1,1})\,,\\[1mm]
\hat e^{u}_2 \!\!\!&=\!\!\!& \mathrm{diag}(e^u_{2,1},0,0,e^u_{2,2},0,e^u_{2,3},0,0,0)\,.
\end{array}
\end{eqnarray}
With this notation, 
and discarding irrelevant operators, which give negligible physical effects 
at low energies, the most general local Lagrangian for the light SM fields and
the heavy flavour messengers that is compatible with the symmetries of the
theory has the structure
\begin{eqnarray}
\mathcal{L} = \mathcal{L}^{\rm bulk}
+ \delta(y) \mathcal{L}^0 
+ \delta(y-\pi R) \mathcal{L}^{\pi R} \,,
\label{Lag}
\end{eqnarray}
with
\begin{eqnarray}
\mathcal{L}^{\rm bulk} \!\!\!&=\!\!\!& 
\sum_{l=u,d} \Big[i \bar{\p}^{l} \gamma^{M} D_{M} \p^{l} 
+ i \bar{\tilde{\p}}{}^{l} \gamma^{M} D_{M} \tilde{\p}^{l} 
-  M_{l} (\bar{\p}^{l} \tilde{\p}^{l} 
+ \bar{\tilde{\p}}{}^{l}\p^{l})\Big]\;, 
\label{lbulk} \\
\mathcal{L}^0 \!\!\!&=\!\!\!&
i \bar{\chi}_L^d \gamma^{\mu} D_{\mu} \chi_L^d 
+ i \bar{\tilde{\chi}}_L^d \gamma^{\mu} D_{\mu} \tilde{\chi}_L^d +
i \bar{\chi}_R^u \gamma^{\mu} D_{\mu} \chi_R^u 
+ i \bar{\tilde{\chi}}_R^u \gamma^{\mu} D_{\mu} \tilde{\chi}_R^u 
\nonumber \\ 
\!\!\!&\;\!\!\!& 
+\,\Big[\bar{\chi}_L^d\, \hat e^{d}_{1}{}^\dagger \,\p^{d} 
+ \bar{\tilde{\chi}}_L^d\, \hat e^{d}_{1}{}^\dagger \,\tilde{\p}^{d}
+ \bar{\chi}^{u}_R\, \hat e^{u}_{1}{}^\dagger \, \p^{u} 
+ \bar{\tilde{\chi}}{}^{u}_R\, \hat e^{u}_{1}{}^\dagger \,\tilde{\p}^{u} 
+ \rm{h.c.}\Big] \,, 
\label{locale1} \\
\mathcal{L}^{\pi R} \!\!\!&=\!\!\!&
i \bar{\chi}^{u}_L \gamma^{\mu} D_{\mu} \chi^{u}_L 
+ i \bar{\tilde{\chi}}{}^{u}_L \gamma^{\mu} D_{\mu} \tilde{\chi}^{u}_L 
+ i \bar{\tilde{\chi}}^d_R \gamma^{\mu} D_{\mu} \tilde{\chi}^d_R 
+ i \bar{\chi}^d_R \gamma^{\mu} D_{\mu} \chi^d_R 
\nonumber \\ 
\!\!\!&\;\!\!\!& 
+\, \Big[\bar{\chi}^d_R\,\hat e^{d}_{2}{}^\dagger\,\p^{d} 
+ \bar{\tilde{\chi}}^d_R\,\hat e^{d}_{2}{}^\dagger\,\tilde{\p}^{d}
+ \bar{\chi}^{u}_L\, \hat  e^{u}_{2}{}^\dagger\,\p^{u} 
+ \bar{\tilde{\chi}}{}^{u}_L\, \hat e^{u}_{2}{}^\dagger \,\tilde{\p}^{u} 
+ \rm{h.c.}\Big] \,. 
\label{locale2}
\end{eqnarray}
We have here tacitly excluded the possibility that odd operators 
might appear in the Lagrangian with coefficients that are themselves 
odd functions of the coordinates, behaving as constants in the bulk 
and jumping discontinuously at the branes. This is reasonable, since 
this kind of odd operators can be distinguished from ordinary even 
operators by a local parity symmetry \cite{localsymmetry}. It should be 
noticed, however, that imposing the latter symmetry significantly 
restricts the possibilities for canceling potential localized anomalies, 
since it forbids bulk Chern-Simons counterterms. 

\subsection{Structure of the induced couplings}
\label{structure}

\begin{figure}[h]
\begin{center} 
\begin{picture}(240,170)(0,-10)
\Line(0,0)(0,100)
\Line(0,0)(40,40)
\Line(40,40)(40,140)
\Line(0,100)(40,140)
\Line(200,0)(200,100)
\Line(200,0)(240,40)
\Line(240,40)(240,140)
\Line(200,100)(240,140)
\Vertex(20,70){1}
\Vertex(220,70){1}
\Line(20,70)(200,70)
\DashLine(200,70)(220,70){3}
\Vertex(120,70){1}
\DashLine(120,70)(120,100){3}
\put(15,80){$Q_L$}
\put(215,80){$q_R$}
\put(117,110){$\alpha$}
\put(67,55){$\Psi_{\bf 2}$}
\put(167,55){$\Psi_{\bf 1}$}
\end{picture}
\caption{\em \small Diagram inducing the effective mass in the presence of $SU(3)_W$ gauge
symmetry breaking only: all the fields carry the same flavour charge. The insertion of $\alpha$
switches from the doublet to the singlet components of the bulk field. Here $Q_L$ and 
$q_R$ can be any left- and right-handed brane fermion and $\Psi$ represents the pair of 
bulk fermions.}
\vskip -10pt
\end{center}
\label{flavdiag}
\end{figure}
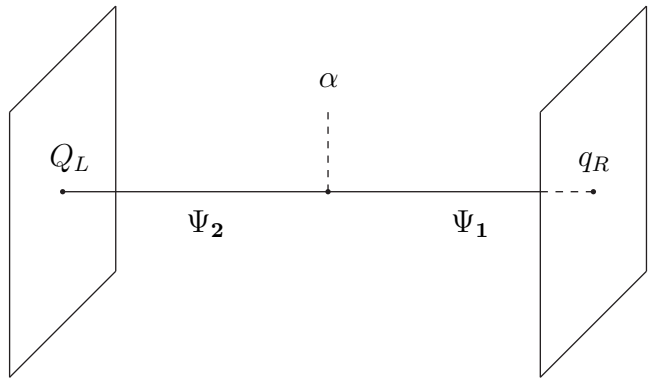

As in ref.~\cite{sss}, effective Yukawa couplings, wave-function 
and vertex corrections for the standard matter fermions are generated in the 
low-energy effective theory defined by integrating out the heavy 
messenger fermions. For example, mass terms are obtained from the 
diagrams in Fig.~1. In this case, however, a given brane fermion 
can couple only to the flavour component of the bulk fermions that 
has the same $U(1)_F$ charge. This implies that a non-vanishing 
Yukawa coupling, wave-function or vertex correction is generated only if
the involved brane fields have equal $U(1)_F$ charge, as long as 
the $U(1)_F$ symmetry stays unbroken, that is for $\beta = 0$. 
The other Yukawa couplings, wave-function and vertex corrections, involving 
brane fields with different $U(1)_F$ charges, can be generated only 
if the $U(1)_F$ symmetry is broken, that is $\beta \neq 0$. In this case, we
have the diagrams in Fig.~2. Since $T_F^1 = (T_F^+ + T_F^-)/2$ can 
change the $U(1)_F$ charge by $1$ unit, in order to connect two brane 
fields with charges differing by some integer $k$, we need $|k|$ 
insertions of $\beta T_F^1$. The effect will thus be of order $\beta^{|k|}$. 

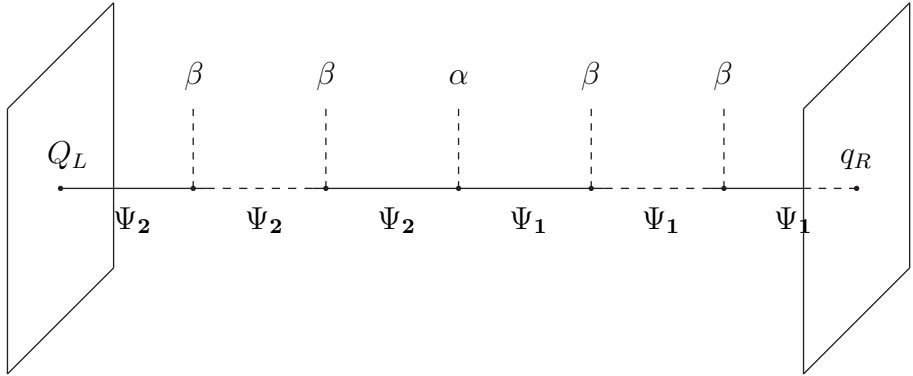
\begin{figure}[h]
\begin{center} 
\begin{picture}(340,170)(0,-10)
\Line(0,0)(0,100)
\Line(0,0)(40,40)
\Line(40,40)(40,140)
\Line(0,100)(40,140)
\Line(300,0)(300,100)
\Line(300,0)(340,40)
\Line(340,40)(340,140)
\Line(300,100)(340,140)
\Vertex(20,70){1}
\Vertex(320,70){1}
\Line(20,70)(75,70)
\DashLine(75,70)(115,70){3}
\Line(115,70)(225,70)
\DashLine(225,70)(265,70){3}
\Line(265,70)(300,70)
\DashLine(300,70)(320,70){3}
\put(15,80){$Q_L$}
\put(315,80){$q_R$}
\Vertex(170,70){1}
\DashLine(170,70)(170,100){3}
\put(167,110){$\alpha$}
\Vertex(70,70){1}
\DashLine(70,70)(70,100){3}
\put(67,110){$\beta$}
\Vertex(120,70){1}
\DashLine(120,70)(120,100){3}
\put(117,110){$\beta$}
\Vertex(220,70){1}
\DashLine(220,70)(220,100){3}
\put(217,110){$\beta$}
\Vertex(270,70){1}
\DashLine(270,70)(270,100){3}
\put(267,110){$\beta$}
\put(40,55){$\Psi_{\bf 2}$}
\put(90,55){$\Psi_{\bf 2}$}
\put(140,55){$\Psi_{\bf 2}$}
\put(190,55){$\Psi_{\bf 1}$}
\put(240,55){$\Psi_{\bf 1}$}
\put(290,55){$\Psi_{\bf 1}$}
\end{picture}
\caption{\em \small Diagram inducing the effective mass in the presence of 
both gauge and flavour symmetry breaking. Each insertion of $\beta$ switches 
between two components of bulk fields with flavour charges differing by one unit. 
The minimal number of such insertions that is needed to get a non-vanishing result
is equal to the difference between the flavour charges of the left- and right-handed
brane fields. Moreover, if this number is even, there is no mass insertion for the bulk 
fields, whereas when it is odd, there must be one mass insertion. Here $Q_L$ and 
$q_R$ can be any left- and right-handed brane fermion and $\Psi$ represents the pair 
of bulk fermions.}
\vskip -10pt
\end{center}
\label{flav}
\end{figure}

Actually, a further restriction turns out to be present, depending on whether 
$k$ is even or odd, as a consequence of the fact that the two types of bulk 
fermions $\psi^l$ and $\tilde \psi^l$ can couple only to SM fermions with 
even and odd flavour charges respectively (in the example under consideration).
The Yukawa couplings can be generated through the exchange of 
bulk fermions with an even or odd number of bulk mass insertions, \textit{i.e.}
with or without a $\psi^l \Leftrightarrow \tilde \psi^l$ transition.
Non-vanishing entries can therefore be generated only with even 
or odd $k$, depending on whether the involved fields couple to the same or 
to a different kind of bulk fields $\p$ or $\tilde{\p}$. 
Wave-function and vertex corrections can instead be generated only with 
an even number of bulk mass insertions, \textit{i.e.} without an overall $\psi^l
\Leftrightarrow \tilde \psi^l$ transition, and a non-vanishing correction is 
therefore generated only for $k$ even.

The above reasoning shows that with a suitable assignment of the 
$SU(2)_F$ quantum numbers for brane and bulk fermions, it is possible 
to induce effective mass matrices with a pattern of matrix elements that can 
naturally explain the hierarchies among observed masses and mixing angles
for matter fermions. Just as with the FN mechanism, the entries of the 
Yukawa couplings $Y^{u,d}_{IJ}$ (from now on we denote by $I$ and $J$ the
family index) and the wave-function factors 
$Z^Q_{IJ}$ and $Z^{u,d}_{IJ}$ for doublets and singlets respectively, can be
expressed as powers of the 
order parameter $\lambda \equiv \pi \beta$ for the breaking of the Abelian
flavour symmetry,
modulo numerical factors of order one. The results can be written in terms 
of the charges $q_I$ of the left-handed doublets $Q$ and the charges 
$l_I$ of the right-handed singlets $l=u,d$ as:
\begin{eqnarray}
\label{expyukawa}
Y^l_{IJ} \!\!\!&\sim\!\!\!& \lambda^{|q_I - l_J|} \,, \\[4mm]
Z^Q_{IJ} \!\!\!&\sim\!\!\!& \left\lbrace\begin{array}{l}
\delta_{IJ} + \lambda^{|q_I - q_J|}\,\quad \mathrm{for} \; |q_I - q_J|\;
\mathrm{even} \\[1mm]
\delta_{IJ}\,\quad \mathrm{for} \; |q_I - q_J|\; \mathrm{odd}
\end{array} \right. \,, \\[1mm]
Z^l_{IJ} \!\!\!&\sim\!\!\!& \left\lbrace \begin{array}{l}
\delta_{IJ} + \lambda^{|l_I - l_J|} \,\quad\mathrm{for} \; |l_I - l_J|\;
\mathrm{even} \\[1mm]
\delta_{IJ} \,\quad\mathrm{for} \; |l_I - l_J|\; \mathrm{odd}
\end{array} \right. \;.
\end{eqnarray}
The physical Yukawa couplings are obtained after performing a transformation 
on matter fermions that brings their kinetic terms to a canonical form.
To do so (see for example \cite{u1sapore}), we first diagonalize the wave functions
as $Z^Q = U^Q{}^\dagger D^Q
U^Q$ and $Z^l = U^l{}^\dagger D^l U^l$ in terms of some unitary
matrices $U^Q$ and $U^l$. In general, the diagonal matrices have 
entries of order one, $D^Q_{II} \sim 1$ and $D^l_{II} \sim 1$, 
but differ from the identity, while the $U$ matrices have the same form as the 
wave-function corrections themselves, \textit{i.e.} $U^Q_{IJ} \sim \delta_{IJ}+
\lambda^{|q_I
- q_J|}$ and $U^l_{IJ} \sim \delta_{IJ}+\lambda^{|l_I - l_J|}$. We then redefine
the matter 
fields to be $\hat Q = \sqrt{D^Q} U^Q Q$ and $\hat l = \sqrt{D^l} U^l
l$. In this way, the new wave-function factors are $\hat Z^q = 1$ and $\hat Z^l
= 1$, 
whereas the new Yukawa coupling is given by $\hat Y^l = (D^Q)^{-\frac 12}
U^Q Y^l U^l{}^\dagger (D^l)^{-\frac 12}$. In terms of $\lambda$ this means
\begin{eqnarray}
\hat Y^l_{IJ} \sim \sum_{KL} \lambda^{|q_I - q_K|+|q_K - l_L|+|l_L - l_J|} 
\sim \lambda^{|q_I - l_J|} \,.
\label{expyukawabis}
\end{eqnarray}
The last step, which follows from the inequality $|x| + |y| \ge |x+y|$, 
shows that as in standard 4D flavour models the wave function corrections 
do not mess up the structure of the Yukawa couplings.

Equations (\ref{expyukawabis}) realize the starting point 
for building interesting flavour models. However, a more careful analysis 
shows that our 5D construction presents a number of peculiarities that 
make it much more constrained than a generic 4D flavour model of the 
FN type, mostly due to the embedding in a non-Abelian group and to the 
structure of the mediator sector:

\begin{itemize} 

\item 
The flavour charge is quantized and charge differences are integer.

\item
The precise numerical factors appearing in the induced Yukawa couplings 
are correlated, and contain potentially large group-theoretical
coefficients.

\end{itemize}

\subsection{Effective Lagrangian and induced couplings}
\label{sec:effective}

We now present the explicit computation of the 4D effective Lagrangian,
and in particular the corrections to the kinetic and mass terms for
the SM fields. The leading effects are
obtained by integrating out the heavy bulk fermions at the classical level.
The computation can be done along the lines of ref.~\cite{sss}. 
In order to illustrate the procedure, we  start by discussing in detail the
$d$-quark sector. The up quark sector will then be
easily explained.

\subsubsection{Mode decomposition}
\label{modedecom}

In general, matter fields obey the compactification conditions in
eqs.~(\ref{eq:orbi}) and (\ref{eq:twist}).
In the following, $SU(3)_{W}$ and $SU(2)_{F}$ indices will be denoted by
$i,j,\dots$ and $a,b,\dots$ respectively, and we work in a basis where
the orbifold projection is diagonal, whereas, in general, the twist is
non diagonal.  For fixed electroweak and flavour indices, the
$\gamma_{5}$ matrix acting in the orbifold projection causes the
right- and left-handed matter field components to have different
parity. Hence we can write the matter field components as follows:
\begin{eqnarray}
\p_{i,a}(x,y)  = \p_{i,a}^{+}(x,y) + \p_{i,a}^{-}(x,y) \;,
\label{partenza}
\end{eqnarray}
where $ \p_{i,a}^{+}(x,y)$ and $ \p_{i,a}^{-}(x,y)$ are fields with
positive and negative orbifold parity respectively and a
given chirality which depends on $i$  and  $a$.  Thus the superscript $\pm$
refers to the orbifold parity. The fields  satisfying the
condition (\ref{eq:orbi}) can be expanded in KK modes as
\begin{eqnarray}
\p_{i,a}^{+}(x,y) \!\!\!&=\!\!\!& \frac{1}{\sqrt{\pi R}} \;\sum_{n=0}^{+
\infty} (\frac{1}{\sqrt{2}})^{\delta_{n,0}} (\p_{i,a}^{+})_n(x)
\; \cos(\frac{n y}{R})\;,
\nonumber \\ 
\p_{i,a}^{-}(x,y) \!\!\!&=\!\!\!&  \frac{1}{\sqrt{\pi R}} \;\sum_{n=1}^{+
\infty}( \p_{i,a}^{-})_n(x) \;\sin(\frac{n y}{R}) \;.
\label{sommaKK}
\end{eqnarray}
It is convenient to express $\p_{i,a}^{+}(x,y)$,
$\p_{i,a}^{-}(x,y)$ as sums over all integer modes, both positive
and negative; this is done by defining the negative modes of a given
component as reflection of the positive modes: 
$(\p_{i,a}^{\pm})\raisebox{-1.5pt}{${}_{-n}^\dagger$}(x)
= \pm (\p_{i,a}^{\pm})\raisebox{-1.5pt}{${}_{n}$}(x)$.
The new mode expansion for untwisted fields is then
\begin{eqnarray}
\p_{i,a}^+(x,y) \!\!\!&=\!\!\!& \frac{1}{\sqrt{2 \pi R}} \;\sum_{n=-
\infty}^{+ \infty} \eta_{n} (\p_{i,a}^+)_{n}(x)\cos(\frac{n y}{R})\;,
\nonumber \\  
\p_{i,a}^-(x,y) \!\!\!&=\!\!\!& \frac{1}{\sqrt{2 \pi R}} \;\sum_{n=-
\infty}^{+ \infty} \eta_{n}(\p_{i,a}^-)_{n}(x)\sin(\frac{n y}{R}) \;, 
\label{sommaKKK}
\end{eqnarray}
where
\begin{eqnarray}
\eta_{n} = \left\{
\begin{array}{lll}
1/\sqrt{2} &\mbox{ if }& n \neq 0 \\[1mm]
1 &\mbox{ if }& n = 0
\end{array}
\right.\,.
\end{eqnarray}
We now  switch to the basis in which the SS twist is
diagonal.  The eigenvectors $\Psi^{\pm}_{i,a}$ of the 
twist $T_{{\cal R},{\cal R}^\prime}$ can be written as follows:
\begin{eqnarray}
\Psi^+_{i,a} =  (U_W)_{ij}^{{\cal R}} (V_F)_{ab}^{{\cal R}^\prime} \p_{j,b}^{+}
\;,\qquad
\Psi^-_{i,a} = (U_W)_{ij}^{{\cal R}} (V_F)_{ab}^{{\cal R}^\prime} \p_{j,b}^{-} 
\;, 
\label{ridefinizioni}
\end{eqnarray}
where $(U_W)^{\cal R}$ and $(V_F)^{{\cal R}^\prime}$ are two unitary matrices
in the gauge and flavour space and the labels ${\cal R}$ and 
${\cal R}^\prime$ denote the representation to which the matter fields
belong.  Since the rotation mixes different  indices $i$ and $a$, 
corresponding to different chiralities, the fields $\Psi_{i,a}^{+}$ and $
\Psi_{i,a}^{-}$ do not have a definite chirality, when expanded 
in KK modes as in eqs.~(\ref{sommaKKK}). On the other hand, since the twist
mixes fields with the same orbifold parity, it is possible to diagonalize it
with transformations acting separately on $\p^+$ and $\p^-$.

In the new basis, the twist is diagonal. 
The explicit expressions for the unitary matrices $(U_W)^{\bf{3}}$, 
$(U_W)^{\bf{6}}$ and $(V_F)^{\bf{9}}$ that diagonalize the twist matrices 
$(T^{6}_W)^{\bf{3}}$, $(T^{6}_W)^{\bf{6}}$ and $(T^{1}_W)^{\bf{9}}$
to the forms $(t_W)^{\bf{3}} = {\rm diag}(1/2,0,-1/2)$, 
$(t_W)^{\bf{6}} ={\rm diag}(1,1/2,0,0,0,-1/2,-1)$ and
$(t_F)^{\bf{9}} ={\rm diag}(-4,4,-3,3,-2,2,-1,1,0)$ are given by:
\begin{eqnarray}
(U_W)^{\bf{3}} \!\!\!&=\!\!\!& \frac{1}{\sqrt{2}}\left(
\begin{array}{ccc}
0 & 1 & 1 \\[1mm]
\sqrt{2} & 0 & 0 \\[1mm]
0 & \mbox{-}1 & 1
\end{array}
\right) \,,\;\;
(U_W)^{\bf{6}} =  \frac{1}{2}\left(
\begin{array}{cccccc}
0 & 0 & 0 & 1 & \sqrt{2} & 1 \\[1mm]
0 & \sqrt{2} & \sqrt{2} & 0 & 0 & 0 \\[1mm]
0 & 0 & 0 & \mbox{-}\sqrt{2} & 0 & \sqrt{2} \\[1mm]
2 & 0 & 0 & 0 & 0 & 0 \\[1mm]
0 & \mbox{-}\sqrt{2} & \sqrt{2} & 0 & 0 & 0 \\[1mm]
0 & 0 & 0 & 1 & \mbox{-}\sqrt{2} & 1
\end{array}
\right) \,, \nonumber \\  
(V_F)^{\bf{9}} \!\!\!&=\!\!\!& \frac{1}{16} \left(
\begin{array}{ccccccccc}
1& \mbox{-} \sqrt{8} & \sqrt{28} & \mbox{-} \sqrt{56} & \sqrt{70} & \mbox{-} \sqrt{56} & \sqrt{28} & \mbox{-} \sqrt{8} & 1 \\[1mm]
1& \sqrt{8} & \sqrt{28} & \sqrt{56} & \sqrt{70} & \sqrt{56} & \sqrt{28} & \sqrt{8} & 1 \\[1mm]
\mbox{-} \sqrt{8} & 6 & \mbox{-} \sqrt{56} & \sqrt{28} & 0 & \mbox{-} \sqrt{28} & \sqrt{56} & \mbox{-} 6 & \sqrt{8} \\[1mm]
\mbox{-} \sqrt{8} & \mbox{-} 6 & \mbox{-} \sqrt{56} & \mbox{-} \sqrt{28} & 0 & \sqrt{28} & \sqrt{56} & 6 & \sqrt{8} \\[1mm]
\sqrt{28} & \mbox{-} \sqrt{56} & 4 & \sqrt{8} & \mbox{-} \sqrt{40} & \sqrt{8} & 4 & \mbox{-} \sqrt{56} & \sqrt{28} \\[1mm]
\sqrt{28} & \sqrt{56} & 4 & \mbox{-} \sqrt{8} & \mbox{-} \sqrt{40} & \mbox{-}\sqrt{8} & 4 & \sqrt{56} & \sqrt{28} \\[1mm]
\mbox{-} \sqrt{56} & \sqrt{28} & \sqrt{8} & \mbox{-} 6 & 0 & 6 & \mbox{-} \sqrt{8} & \mbox{-} \sqrt{28} & \sqrt{56} \\[1mm]
\mbox{-} \sqrt{56} & \mbox{-} \sqrt{28} & \sqrt{8} & 6 & 0 & \mbox{-} 6 & \mbox{-} \sqrt{8} & \sqrt{28} & \sqrt{56} \\[1mm]
\sqrt{70} & 0 & \mbox{-} \sqrt{40} & 0 & 6 & 0 & \mbox{-} \sqrt{40} & 0 & \sqrt{70}
\end{array}
\right) \,.\qquad
\end{eqnarray}
Let us now define 
\begin{eqnarray}
\hat\Psi_{i,a} = \left( \Psi_{i,a}^{+},
\Psi_{i,a}^{-}\right) \;.
\end{eqnarray}
In this basis, the effect of the twist amounts to shifting the masses of 
the KK modes by the quantity $2(t_W)_{ii} \alpha + 2(t_F)_{aa}
\beta$.
Therefore, suppressing all the indices, the new KK mass spectrum is given by
\begin{eqnarray}
m_n(\alpha,\beta) = \frac {n \sigma_1 + (2 t_W \alpha + 2 t_F \beta)
1\!\!1}R \,,
\label{massspectrum0}
\end{eqnarray}
where $\sigma_1$ and $1\!\!1$ act at fixed $i,a$ on the space 
$(\Psi_{i,a}^{+}, \Psi_{i,a}^{-})$ and connect terms
with opposite and equal orbifold parity respectively.
Finally, a complete diagonalization can be achieved by mixing states with opposite 
orbifold chirality:
\begin{equation}
(\Psi_{i,a})_n=\eta_n\left[(\Psi_{i,a}^{+})_n+ (\Psi_{i,a}^{-})_n\right]\,
\end{equation}   
where now positive and negative $n$ components of $\Psi_{i,a}$ are independent, and 
their mass is simply given by
 \begin{eqnarray}
m_n(\alpha,\beta) = \frac {n + (2 t_W \alpha + 2 t_F \beta)}R \,.
\label{massspectrum}
\end{eqnarray}

\subsubsection{Construction of the effective Lagrangian}
\label{downq}

In order to derive the effective Lagrangian that is induced for 
the SM fermions by integrating out the bulk fermions at the classical 
level, we use for the latter the mode decomposition derived in previous 
subsection, and switch to 4D momentum space. The relevant linear and 
quadratic parts of the Lagrangian for the modes of the bulk fermions then 
becomes
\begin{eqnarray}
\mathcal{L} = \sum_{n=-\infty}^{\infty} 
\Big[\mathcal{L}^{\rm bulk}_n + \mathcal{L}^{0}_n 
+ (-1)^n \mathcal{L}^{\pi R}_n \Big] \,,
\label{Lagn}
\end{eqnarray}
where
\begin{eqnarray}
\mathcal{L}^{\rm bulk}_{n} \!\!\!&=\!\!\!& \sum_{l=u,d} \Big[
\bar{\Psi}^{l}_n \big(p\!\!\!/ - m_n \big) \Psi^{l}_n 
+ \bar{\tilde \Psi}{}^{l}_n \big(p\!\!\!/ + m_n \big) \tilde \Psi^{l}_n 
- M_l \Big(\bar{\Psi}^{l}_n \tilde{\Psi}^{l}_n +
\bar{\tilde{\Psi}}{}^{l}_n \Psi^{l}_n \Big) \Big]\,,
\label{bulkn} \\ 
\mathcal{L}^{0}_n \!\!\!&=\!\!\!& 
\frac{1}{\sqrt{2 \pi R}} \Big[
\bar{\chi}^d_L (U_W V_F\, \hat e^{d}_{1})^\dagger  \Psi^{d}_n 
+ \bar{\tilde{\chi}}^d_L (U_W V_F\, \hat e^{d}_{1})^\dagger \tilde{\Psi}^{d}_n 
\nonumber \\ \!\!\!&\;\!\!\!&  \hspace{35pt} 
+\, \bar{\chi}^{u}_R (U_W V_F\, \hat e^{u}_{1})^\dagger \Psi^{u}_n 
+ \bar{\tilde{\chi}}{}^{u}_R (U_W V_F\, \hat e^{u}_{1})^\dagger
\tilde{\Psi}^{u}_n 
+ \rm{h.c.}\Big] \,, 
\label{locale1n} \\
\mathcal{L}^{\pi R}_n \!\!\!&=\!\!\!&
\frac{1}{\sqrt{2 \pi R}}\Big[
\bar{\chi}^d_R (U_W V_F\, \hat e^{d}_{2})^\dagger \Psi^{d}_n
+ \bar{\tilde{\chi}}^d_R (U_W V_F\, \hat e^{d}_{2})^\dagger \tilde{\Psi}^{d}_n
\nonumber \\ \!\!\!&\;\!\!\!&  \hspace{35pt} 
+ \bar{\chi}^{u}_L (U_W V_F\, \hat e^{u}_{2})^\dagger \Psi^{u}_n 
+ \bar{\tilde{\chi}}{}^{u}_L (U_W V_F\, \hat e^{u}_{2})^\dagger
\tilde{\Psi}^{u}_n 
+ \rm{h.c.}\Big] \,. 
\label{locale2n}
\end{eqnarray}
From these expressions it is clear that the physics of the light 
modes depends on the mass mixings $\hat e_{1,2}/\sqrt{2 \pi R}$ encoding 
the couplings between brane and bulk modes and on the masses $M_l$
for the bulk modes. The relevant dimensionless parameters are then 
the products of these masses with the length $\pi R$ of the internal 
dimension:
\begin{eqnarray}
\epsilon_{1,2}^l = \sqrt{\pi R/2}\, \hat e_{1,2}^l \,,\qquad
x_l = \pi R M_l \,.
\label{par}
\end{eqnarray}
For convenience, we also define the $\epsilon$ couplings in the basis of
diagonal twist:
\begin{eqnarray}
\varepsilon_{1,2}^l = U_W V_F \epsilon_{1,2}^l \,.
\label{pardiag}
\end{eqnarray}

The mass and wave-function corrections are generated by 
diagrams similar to the ones in Figs.~1 and 2. The result depends 
on the vacuum expectation value of the 
Higgs field $A_5$ through the dimensionless parameter 
$\alpha= g_5 R \langle A_5\rangle/2$, and on the phase $\beta$ induced
by the twist. If the flavour symmetry were local, $\beta$ would be
related to the fifth component of the corresponding $SU(2)_F$ field.
For a global symmetry, $\beta$ is a free parameter related to the
phase accumulated by the twist. The tree-level
propagator in momentum space for the KK modes of $\Psi^{l}$ and
$\tilde{\Psi}^{l}$ is given, in two-by-two matrix notation, by
the following expression:
\begin{eqnarray}
S_l^{n} = \frac{i }{p^{2} - m_n(\alpha,\beta)^{2} 
- (M_l)^2}\left( 
\begin{array}{cc}
 p\!\!\!/ + m_n(\alpha,\beta) &  M_l\\[1mm]
 M_l & p\!\!\!/ - m_n(\alpha,\beta) 
\end{array}
\right). 
\label{propagatorebulk1}
\end{eqnarray}
The effective action is then obtained by integrating out the bulk fermions at the classical 
level, using the above propagator and treating the brane fields as sources, as in ref.~\cite{sss}.
We find a contribution to the effective Lagrangian in momentum space of the form
$\mathcal{L}^{\rm eff}_{d} = \mathcal{L}^{\rm kin}_{d} + \mathcal{L}^{\rm m}_{d}$, 
where $\mathcal{L}^{\rm kin}_{d}$ contains the kinetic term corrections of SM matter fields 
and is given by
\begin{eqnarray}
\mathcal{L}^{\rm kin}_{d} \!\!\!&=\!\!\!& 
\;\overline{\chi}_L^d \varepsilon_{1}^{d^\dagger} \;(p\!\!\!/)\;
F(p,M_{d},2 t_W  \alpha + 2 t_F \beta) 
\;\varepsilon_{1}^{d } \chi_L  \nonumber \\[1mm] 
\!\!\!&\;\!\!\!& + \;\overline{\chi}_R^d \varepsilon_{2}^{d^\dagger}
\;(p\!\!\!/) \;F(p,M_{d},2 t_W  \alpha + 2 t_F\beta) \;
\varepsilon_{2}^{d } \chi_R^d  \nonumber \\[1mm] 
\!\!\!&\;\!\!\!& + \;\overline{\tilde{\chi}}_L^d  \varepsilon_{1}^{d^\dagger} 
\;(p\!\!\!/ ) \;F(p,M_{d}, 2 t_W  \alpha + 2 t_F \beta) \;
\varepsilon_{1}^{d } \tilde{\chi}_L^d \nonumber \\[1mm] 
\!\!\!&\;\!\!\!& + \;\overline{\tilde{\chi}}_R^d \varepsilon_{2}^{d^\dagger} 
\;(p\!\!\!/)  \;F(p,M_{d}, 2 t_W \alpha + 2 t_F \beta) \; 
\varepsilon_{2}^{d }\tilde{\chi}_R^d \;, 
\label{wfc}
\end{eqnarray}
whereas $\mathcal{L}^{\rm m}_{d}$ contains the effective mass terms
and is given by
\begin{eqnarray}
\mathcal{L}^{\rm m}_{d} \!\!\!&=\!\!\!& \frac{1}{\pi R}
\bigg\{ \left[  \overline{\chi}_L^d  \varepsilon_{1}^{d^\dagger} 
\;G_{1}(p,M_{d}, 2 t_W  \alpha + 2 t_F \beta)  \;
\varepsilon_{2}^{d} \chi_R^d  + 
\mathrm{h.c.} \right] \nonumber \\ 
\!\!\!&\;\!\!\!& \hspace{23pt} -  \left[ \overline{\tilde{\chi}}_L^d  \varepsilon_{1}^{d^\dagger} 
\;G_{1}(p,M_{d}, 2 t_W  \alpha + 2 t_F \beta)  \;
\varepsilon_{2}^{d}\tilde{\chi}_R^d + 
\mathrm{h.c.} \right] \nonumber \\[1mm] 
\!\!\!&\;\!\!\!& \hspace{23pt} + \left[ \overline{\chi}_L^d \varepsilon_{1}^{d^\dagger} \;
G_{2}(p,M_{d}, 2 t_W  \alpha + 2 t_F \beta) \;
\varepsilon_{2}^{d}\tilde{\chi}_R^d  + \mathrm{h.c.}\right] \nonumber \\
\!\!\!&\;\!\!\!& \hspace{23pt} + \left[ \overline{\tilde{\chi}}_L^d  \varepsilon_{1}^{d^\dagger} \;
G_{2}(p,M_{d}, 2 t_W  \alpha +2 t_F \beta) \;  \varepsilon_{2}^{d}
\chi_R^d + \mathrm{h.c.}\right] \bigg\} \;.  
\label{massterm}
\end{eqnarray}
For the $u$ quarks, one can proceed exactly in the same way. 
In  the Euclidean space-time,  the explicit expressions of the
functions $F$, $G_{1}$ and $G_{2}$  are given by 
\begin{eqnarray}
F(p,M,\rho)  \!\!\!&=\!\!\!& \frac{1}{(\pi R)^{2}}
\sum_{n=-\infty}^{\infty} \frac{1}{p^{2} +
(\frac{n + \rho}R)^{2} + M^2} \nonumber \\ 
\!\!\!&=\!\!\!& \frac{1}{\pi R \sqrt{p^{2} + M^{2}}}
\;\mathrm{Re}\left[\sum_{m=-\infty}^{\infty} e^{- |2m|  \pi R \sqrt{p^{2} +
M^2}} e^{- |2 m| \pi i \rho} \right] \nonumber \\ 
\!\!\!&=\!\!\!& \frac{1}{\pi R \sqrt{p^{2} + M^{2}}}
\;\mathrm{Re}\coth(\pi R \sqrt{p^{2} + M^{2}}+ i \rho \pi)\,, \nonumber \\ 
G_{1}(p,M,\rho)  \!\!\!&=\!\!\!& \frac{1}{\pi R}
\sum_{n=-\infty}^{\infty} (-1)^{n}
\frac{\frac{n + \rho}{R}}{p^{2} + (\frac{n
+ \rho}R)^{2} + M^{2}} \nonumber \\ 
\!\!\!&=\!\!\!& \mathrm{Im}\left[\sum_{m=-\infty}^{\infty} e^{- |2m+1|
\pi R \sqrt{p^{2} + M^{2}}} e^{- |2 m +1|\pi i \rho }\right] \label{derf} \\ 
\!\!\!&=\!\!\!& - \mathrm{Im}\csch(\pi R \sqrt{p^{2} + M^{2}}+i \rho \pi)\,, \nonumber \\[2mm]
G_{2}(p,M,\rho)  \!\!\!&=\!\!\!& \frac{1}{\pi R}
\sum_{n=-\infty}^{\infty}(-1)^{n}
\frac{M}{p^{2} + (\frac{n + \rho}{R})^{2} + M^{2}} \nonumber \\ 
\!\!\!&=\!\!\!& \frac{M}{\sqrt{p^{2} + M^{2}}} \;
\mathrm{Re}\left[\sum_{m=-\infty}^{\infty} e^{- |2m+1| 
\pi R \sqrt{p^{2} + M^{2}}} 
e^{- |2 m+1| \pi i \rho} \right]\nonumber \\
\!\!\!&=\!\!\!& \mathrm{Re}\csch(\pi R \sqrt{p^{2} + M^{2}}+i \rho \pi)\,. 
\nonumber
\end{eqnarray}

\subsection{Low energy limit}
\label{elale}

We now study the effective Lagrangian in the low energy 
limit $p^{2} \ll M^{2}_l$. In this limit, the non-local $p$-dependent 
couplings of eqs.~(\ref{wfc})-(\ref{massterm}) and the analogous terms for
up-type quarks reduce to local kinetic 
and mass terms. After diagonalization and canonical normalization of 
the physical fields, these generate the 
physical fermion masses and mixings.

In the low-energy limit $p^{2} \ll M^{2}_l$, the momentum variable 
$\pi R \sqrt{p^{2} + M^{2}_l}$ reduces to the constant parameter $x_l$ 
defined in eq.~(\ref{par}). The functions $F$, $G_1$ and $G_2$ become 
simple trigonometric functions of the three parameters $x_l$, $\alpha$ and $\beta$.
Notice moreover that not all the functional dependence on the parameters 
$\alpha$ and $\beta$ is relevant. First of all, for various phenomenological 
reasons that were explained in ref.~\cite{sss} and in sec.~\ref{ghu} and that 
we will review below, we must assume that $\alpha$ is small and retain 
only the leading effects that are at most linear in $\alpha$. Since $\alpha$ is 
related to the VEV of the Higgs field, this corresponds to keeping only those
effective operators that involve at most one Higgs field. Moreover, it is easy 
to check that only even powers of $\beta$ are relevant in $F$ and $G_1$, and 
similarly only odd powers of $\beta$ are relevant in $G_2$, due to the flavour 
quantum numbers of the brane fields (see eqs.~(\ref{chid})-(\ref{chiut})). 
The above functions can therefore be effectively substituted with:
\begin{eqnarray}
F(p,M_l,2 t_W\alpha+2 t_F\beta) \!\!\!&\Rightarrow\!\!\!& f(x_l,2 t_F \beta)
\,, \nonumber \\ 
G_{1}(p,M_l,2 t_W\alpha+2 t_F\beta) \!\!\!&\Rightarrow\!\!\!& (2 \pi t_W
\alpha)\,
g_1(x_l,2 t_F \beta) \,, 
\label{fggapprossimate} \\ 
G_{2}(p,M_l,2 t_W\alpha+2 t_F\beta) \!\!\!&\Rightarrow\!\!\!& (2 \pi t_W
\alpha)\,
g_2(x_l,2 t_F \beta) \,, \nonumber 
\end{eqnarray}
where
\begin{eqnarray}
f(x_l,2 t_F \beta) \!\!\!&=\!\!\!& \frac{1}{x_l}\, 
\mathrm{Re}\,\coth\Big[x_l + 2 \pi i t_F \beta\Big] \,, \nonumber \\
g_1(x_l,2 t_F \beta) \!\!\!&=\!\!\!& 
\mathrm{Re}\,\left(\coth \Big[x_l + 2 \pi i t_F \beta \Big] 
\csch \Big[x_l + 2 \pi i t_F \beta \Big]\right) \,, \label{fggapprossimatebis}\\
g_2(x_l,2 t_F \beta) \!\!\!&=\!\!\!& 
\mathrm{Im}\,\left(\coth \Big[x_l + 2 \pi i t_F \beta \Big] 
\csch \Big[x_l + 2 \pi i t_F \beta \Big]\right) \raisebox{16pt}{}\,. \nonumber
\end{eqnarray}

Let us be more quantitative on the range of values that the above 
dimensionless parameters are allowed to take by basic phenomenological 
constraints. A first important requirement is that $m_W \ll 1/R$, since 
indirect experimental constraints imply that the compactification scale 
should be at least a few TeV. A second requirement is that $M_l \gg m_W$, 
in such a way that even the lightest modes of the extra bulk fermions 
that we have introduced are heavy enough to satisfy direct experimental 
constraints. This two conditions imply respectively the following 
restrictions:
\begin{eqnarray}
\pi \alpha \ll 1 \,,\qquad 
\pi \alpha \ll x_l \,.
\end{eqnarray}
Notice that the above conditions justify in a more precise way 
the approximation done to derive eqs.~(\ref{fggapprossimatebis}).
Notice also that they do not fix the size of the parameters $x_l$ related to 
the masses of the bulk fermions.  

The total effective Lagrangian is obtained by adding up the first 
rows of the brane Lagrangians (\ref{locale1})-(\ref{locale2}) and 
the correction $\mathcal{L}^{\rm eff}_u + \mathcal{L}^{\rm eff}_d$. After
simplifying the traces over gauge and flavour indices, which in the
approximation leading to eqs.~(\ref{fggapprossimatebis}) are disentangled, it
can be rewritten in terms of the original three generations of fields $u_L$,
$u_R$, $d_L$, $d_R$ and couplings $\epsilon_{L,R}^l$, and has the following 
general form:
\begin{eqnarray} 
{\cal L}^{{\rm phen}} \!\!\!&=\!\!\!&  \sum_{a,b=1}^3 \bigg\{
\bar u_{L}^a \, p\!\!\!/ \, {\cal Z}^{u_L}_{ab} u_{L}^b 
+ \bar u_{R}^a \, p\!\!\!/ \, {\cal Z}^{u_R}_{ab} u_{R}^b 
+ \Big(\bar u_{L}^a \mathcal{M}^{u}_{ab} u_{R}^b + {\rm h.c.} \Big) 
\nonumber \\  \!\!\!&\;\!\!\!& \hspace{28pt} +\,  
\bar d_{L}^a \, p\!\!\!/ \, {\cal Z}^{d_L}_{ab} d_{L}^b 
+ \bar d_{R}^a \, p\!\!\!/ \, {\cal Z}^{d_R}_{ab} d_{R}^b 
+ \left(\bar d_{L}^a \mathcal{M}^{d}_{ab} d_{R}^b + {\rm h.c.}
\right)\bigg\}\,. 
\label{lagrph}
\end{eqnarray}

\subsection{Fermion masses and mixings}
\label{sec:masses}

Let us now specialize to the case at hand and work out in detail
the expressions for fermion masses and mixing angles that can be obtained from
${\cal L}^{{\rm phen}}$ in eq.~(\ref{lagrph}). In order to make the physics
behind ${\cal L}^{{\rm phen}}$ clear, it is instructive to study the two
limits $x_l \ll 1$ and $x_l \gg 1$, where many of the expressions drastically
simplify. We start by discussing the case $x_l \gg 1$, since the corrections to
the quark field wave functions are simpler in this limit.

\subsubsection*{$\mathbf{x_l \gg 1}$}

In the limit of $x_l \gg 1$, the functions in eqs.~(\ref{fggapprossimatebis}) 
take the form
\begin{eqnarray} 
f(x_l,2 t_F \beta) \!\!\!&\sim\!\!\!& \frac{1}{x_l} \left(1+2\, e^{-2 x_l} \cos 4 \pi t_F \beta\right) 
\sim  \frac{1}{x_l}\,,\nonumber \\ 
g_1(x_l,2 t_F \beta) \!\!\!&\sim\!\!\!& 2 e^{-x_l} \cos(2 \pi t_F \beta) \,, \label{eq:xgg1} \\[2mm]
g_2(x_l,2 t_F \beta) \!\!\!&\sim\!\!\!& - 2 e^{-x_l} \sin(2 \pi t_F \beta) \,. \nonumber 
\end{eqnarray}
Under the hypothesis that 
$\lambda=\pi\beta$ is of the order of the Cabibbo angle, we expand up to the
appropriate order equations (\ref{eq:xgg1}). For the case at hand, this order
is $\lambda^8$. Our expansion gives the following effective mass matrices:
\begin{eqnarray}  
\mathcal{M}^{d}_{ad} \!\!\!&=\!\!\!&
-  \;m_W \;e^{- x_{d}} 
\,(\mathcal{E}_{1}^{d})^\dagger_{ab} \,\widetilde{Y}^{d}_{bc}\,
(\mathcal{E}_{2}^{d})_{cd}\;\,\label{mamd} \\[1mm]
\mathcal{M}^{u}_{ad} \!\!\!&=\!\!\!&
- \sqrt{2}  \;m_W \;e^{- x_{u}}
 \,(\mathcal{E}_{1}^{u})^\dagger_{ab}\, \widetilde{Y}^{u}_{bc}\,
(\mathcal{E}_{2}^{u})_{cd} \label{mamu} \;,
\end{eqnarray}
where, keeping only the leading terms for each entry,
\begin{eqnarray}
\widetilde{Y}^{d} \!\!\!&=\!\!\!& \left( 
\begin{array}{ccc}
-2\,{\sqrt{14}}\,\lambda^5 & {\sqrt{70}}\,\lambda^4 &2\,{\sqrt{14}}\,\lambda^3 \\[1mm]
-5\,{\sqrt{7}}\,\lambda^4 & 2\,{\sqrt{35}}\,\lambda^3 & 3\,{\sqrt{7}}\,\lambda^2 \\[1mm]
10\,\lambda^2 & -2\,{\sqrt{5}}\,\lambda & -1
\end{array}
\right) \;, \label{babd} \\
\widetilde{Y}^{u} \!\!\!&=\!\!\!& \left( 
\begin{array}{ccc}
\lambda^8 & -2\,{\sqrt{14}}\,\lambda^5 & 2\,{\sqrt{14}}\,\lambda^3 \\[1mm]
2\,{\sqrt{2}}\,\lambda^7 & -5\,{\sqrt{7}}\,\lambda^4 & 3\,{\sqrt{7}}\,\lambda^2 \\[1mm]
-2\,{\sqrt{14}}\,\lambda^5 & 10\,\lambda^2 & -1
\end{array}
\right)
\;,
\label{babu}
\end{eqnarray}
and for convenience we have defined
\begin{equation}
\label{eq:epsiloni}
\mathcal{E}_{k}^{d} = \mathrm{diag}( \epsilon_{k,1}^{d},  
\epsilon_{k,2}^{d}, \epsilon_{k,3}^{d})\,, \qquad
\mathcal{E}_{k}^{u} = \mathrm{diag}( \epsilon_{k,1}^{u},  
\epsilon_{k,2}^{u},\epsilon_{k,3}^{u})\,.  
\end{equation}
We see from eqs.~(\ref{babd}) and (\ref{babu}) that we have obtained the
desired structure in powers of $\lambda$, but the group-theoretical coefficients
are large and modify substantially masses and mixing angles. However,
since these coefficients are entirely fixed by the flavour symmetry, one can
tolerate their presence and design the texture 
in such a way to obtain suitable additional powers of $\lambda$ to compensate 
for the fact that they are not of order $1$. In other words, we can still obtain a good 
description of masses and mixings in terms of a single parameter $\lambda$, but 
with non-conventional textures, which take into account the fact that the numerical 
coefficients can become of order $\lambda^{-1}$ or larger. 
We will discuss in sec.~\ref{sec:settemezzi} an explicit realization of this idea. 
The wave-function corrections are instead given by
\begin{eqnarray}
\mathcal{Z}^{u_L}  \!\!\!&=\!\!\!&  \mathcal{Z}^{d_L}  = 
\mathbf{1} + \frac{1}{x_d} \mathcal{E}_{1}^{d\dagger}
 \mathcal{E}_{1}^{d}
+ \frac{1}{x_u} \mathcal{E}_{1}^{u\dagger}
 \mathcal{E}_{1}^{u} \,, \nonumber \\ 
\mathcal{Z}^{d_R}  \!\!\!&=\!\!\!&  
\mathbf{1} + \frac{1}{x_d} \mathcal{E}_{2}^{d\dagger}
 \mathcal{E}_{2}^{d} \,, \label{az}  \\
\mathcal{Z}^{u_R}   \!\!\!&=\!\!\!&  
\mathbf{1} + \frac{1}{x_u}  \mathcal{E}_{2}^{u\dagger}
\mathcal{E}_{2}^{u}\,. \nonumber
\end{eqnarray}

The physical quark Yukawa couplings are obtained by redefining the quark 
fields to reabsorb the wave-function corrections $\mathcal{Z}$. The 
structure of the latter is such that the physical mass matrix cannot 
grow indefinitely when the $\epsilon_a^{u,d}$ are increased. The reason 
is that the $\epsilon$-parameters encode the mixing between bulk and brane 
fermions.  The resulting mass of the  hybrid fields must therefore 
interpolate between the value that one would get for a bulk field 
($\epsilon_a^{u,d} \rightarrow \infty$) and the vanishing value that one 
would get for a brane field ($\epsilon_a^{u,d} \rightarrow 0$).

In the simple case where the $\epsilon_a^{u}$ and $\epsilon_a^{d}$ are
real, it is useful to introduce the following bulk-brane mixing angles:
\begin{eqnarray}
\begin{array}{ll}
\alpha_{1,a}^{u} = {\rm Arctan}\left(\displaystyle{\frac 
{\sqrt{1/{x_u}}\,\epsilon_{1,a}^{u}}
{\sqrt{1 + 1/{x_d}\,\left(\epsilon_{1,a}^{d}\right)^2}}}\right) \;,\quad 
&\alpha_{1,a}^{d} = {\rm Arctan}\left(\displaystyle{\frac 
{\sqrt{1/{x_d}}\,\epsilon_{1,a}^{d}}
{\sqrt{1 + 1/{x_u}\,\left(\epsilon_{1,a}^{u}\right)^2}}}\right) \;, \medskip \\
\alpha_{2,a}^{u} = {\rm Arctan}\left(\sqrt{1/{x_u}}\,\epsilon_{2,a}^{u}\right) 
\;,\quad 
&\alpha_{2,a}^{d} = {\rm Arctan}\left(\sqrt{1/{x_d}}\,\epsilon_{2,a}^{d}\right)
\;.
\end{array}
\end{eqnarray}
The physical masses, obtained by rescaling the quarks fields in order
to have a canonically normalized kinetic term, e.g. $ \bar
u_{L} \; p\!\!\!/ \; u_{L}$, are then found to be:
\begin{eqnarray}  
\mathcal{M}^u  \!\!\!&=\!\!\!&  
-\, \sqrt{2}\, x_u\, e^{-x_u}\,  m_W\, 
S_1^u\, \widetilde{Y}^{u}\, S_2^u \,, \nonumber \\
\mathcal{M}^d  \!\!\!&=\!\!\!&  
-\,  x_d\, e^{-x_d}\, m_W\,
S_1^d\, \widetilde{Y}^{d}\, S_2^d\,,
\label{physmass}
\end{eqnarray}
where
\begin{eqnarray}
S_1^l  \!\!\!&=\!\!\!&  \mathrm{diag}(\sin \alpha_{1,1}^{l}, \sin \alpha_{1,2}^{l},  \sin \alpha_{1,3}^{l})\,, \nonumber \\
S_2^{l}  \!\!\!&=\!\!\!& \mathrm{diag}(\sin \alpha_{2,1}^{l}, \sin \alpha_{2,2}^{l},  \sin \alpha_{2,3}^{l})\,.
\label{eq:sines}
\end{eqnarray}
At this point, we proceed exactly as in the SM, by diagonalizing the
mass matrices via a bi-unitary transformation
\begin{eqnarray}
u_{L,R}^\alpha   \rightarrow {\cal U}_{L,R}^{\alpha \beta}
u_{L,R}^\beta\;, \quad  
d_{L,R}^\alpha  \rightarrow {\cal D}_{L,R}^{\alpha \beta}
d_{L,R}^\beta \;\; \Rightarrow \;\;
V_{CKM} = {\cal U}^\dagger_L {\cal D}_L  \,.
\end{eqnarray}
The masses in eq.~(\ref{physmass}) are  suppressed with respect to
$m_W = \alpha/R$ by the factor $x_l e^{-x_l}$, which is a small parameter
since we are now considering the limit $x_l \gg 1$, and by a
trigonometric factor parametrizing the bulk-brane mixing. In this
situation we therefore obtain mass matrices with an absolute scale
much smaller than the W mass:
\begin{equation}  
\mathcal{M}^{u,d} \sim x_{u,d} e^{-x_{u,d}} m_W \ll m_W \,.
\end{equation}  
This is phenomenologically not acceptable for the top quark mass. 
Notice, nevertheless, that the exponential dependence on $x_u$ and 
$x_d$ of the overall scale for the masses in the up and down sectors 
could allow to account for the significant hierarchy observed between the latter 
through a modest hierarchy between the two parameters $x_u$ and $x_d$.
The physical origin of the above exponential suppression is related to
the higher-dimensional gauge symmetry constraining the Higgs
interactions.  More precisely, the only invariant Yukawa-type
effective operators turn out to involve the Higgs field in the form of
a Wilson line, which connects the two branes where the relevant left-
and right-handed fermions are located and winds at least once around
the internal interval \cite{sss}. The exchanged bulk fermion of mass
$M_l$ must therefore propagate at least over a distance $\pi R$ and
this implies a suppression factor proportional to $e^{-x_l}$ in the
limit $x_l \gg 1$.

\subsubsection*{$\mathbf{x_l \ll 1}$}
\label{sec:xll1}

In the limit of $x_l \ll 1$, the functions in eqs.~(\ref{fggapprossimatebis}) 
also simplify. Actually, to have a significant simplification 
we really need $x_l \ll \pi \beta$, but deriving an asymptotic expression
in this limit would contrast with the philosophy of flavour models, which
always assumes a power expansion in the order parameter $\pi \beta \ll 1$.
For this reason, we will consider this situation only for the case of 
flavour-singlet bulk fermions, which are blind to the flavour symmetry. 
We will see in sec.~\ref{sec52} that it is possible to take advantage of the 
possibility of adding such a flavour-neutral fermion, in addition to 
flavour-charged ones, to improve the magnitude of the masses of the 
third familiy of quarks. We therefore set $\pi t_F \beta$ to $0$. Under these 
assumptions, the functions of eqs.~(\ref{fggapprossimatebis}) reduce to
\begin{equation}
f(x_l,0) \simeq \frac{1}{x_l^2} 
\,,\qquad  
g_1(x_l,0) \simeq \frac {1}{x_l^2}
\,,  \qquad
g_2(x_l,0) \simeq 0 \,. 
\label{eq:xll1}
\end{equation}
The induced wave functions are then given by (there is no matrix structure here 
since we are considering flavour singlets):
\begin{eqnarray}  
{\cal Z}_{L}^l \simeq 1 
+ \frac{1}{x_d^2} \epsilon_{L}^{d\dagger} \epsilon_{L}^{d} 
+ \frac{1}{x_u^2} \epsilon_{L}^{u\dagger} \epsilon_{L}^{u} \,,\qquad
{\cal Z}_{R}^{l} \simeq 1 
+ \frac{1}{x_l^2} \epsilon_{R}^{l\dagger} \epsilon_{R}^{l} \,.
\label{Zbis}
\end{eqnarray}
Similarly, the induced masses are found to be
\begin{eqnarray}
\mathcal{M}^{u} \simeq \sqrt{2} \frac 1{x_{u}^2} 
\epsilon_{L}^{u^\dagger} \epsilon_{R}^{u}\,m_W\,, \qquad \mathcal{M}^{d} \simeq
\frac 1{x_{d}^2} 
\epsilon_{L}^{d^\dagger} \epsilon_{R}^{d}\,m_W \,.
\label{massette}
\end{eqnarray}
The physical quark masses emerging after canonical normalization are then 
found to be
\begin{eqnarray}
m^u \simeq \sqrt{2} \sin \alpha_L^u \sin \alpha_R^u\, m_W\,, \qquad m^d \simeq
\sin \alpha_L^d \sin \alpha_R^d\, m_W \,,
\label{massebis}
\end{eqnarray}
where now
\begin{eqnarray}
\begin{array}{ll}
\alpha_{L}^{u} = {\rm arctan}\sqrt{\displaystyle{\frac 
{(\epsilon_{L}^{u})^2/{x_u^2}\raisebox{10pt}{}}
{1 + (\epsilon_{L}^{d})^2/{x_d^2}}}} \,,\quad 
&\alpha_{L}^{d} = {\rm arctan}\sqrt{\displaystyle{\frac 
{(\epsilon_{L}^{d})^2/{x_d^2}}
{1 + (\epsilon_{L}^{u})^2/{x_u^2}}}} \,, \medskip \\
\alpha_{R}^{u} = {\rm arctan}\sqrt{(\epsilon_{R}^{u})^2/{x_u^2}} \,,\quad 
&\alpha_{R}^{d} = {\rm arctan}\sqrt{(\epsilon_{R}^{d})^2/{x_d^2}} \,.
\end{array}
\label{anglesbis}
\end{eqnarray}
In this case the quark masses are of order $m_W$. In this situation we 
can therefore achieve mass matrices with a trivial flavour structure 
but a sizable magnitude:
\begin{equation}  
\frac {m^l}{m_W} \sim 1 \,.
\end{equation}  
Notice also that for $\epsilon_{L,R}^l \sim 1$ the angles (\ref{anglesbis})
parametrizing the brane-bulk mixings tend to the large values 
$\alpha_L^u \simeq \delta$, $\alpha_L^d \simeq \pi/2 -\delta$
and $\alpha_R^l \simeq \pi/2$, with 
$\delta = {\rm Arctan}(\epsilon_L^u/\epsilon_L^d x^d/x^u)$, 
reflecting the fact that since $\epsilon_{L,R}^l \gg x_l$ the brane-bulk 
mixing is maximal; the masses (\ref{massebis}) tend then to 
$m^d \simeq \cos \delta\, m_W$ and 
$m^u \simeq \sqrt{2} \sin \delta\, m_W$.

\subsubsection*{$x_l\sim 1$}

In the general case $x_l\sim 1$, the effect of the wave-function corrections 
on the $\mathcal{O}(1)$ numerical coefficients in the physical Yukawa
couplings depends in a complicated way on the parameters $x_l$ and $\epsilon_{1,2}^l$,
and must be separately studied for each point in this parameter space. 
We will present the results of this general analysis in sec.~\ref{minimal}.
It is however clear that the induced masses will always have a
scale that is parametrically given by $m_W$ times some suppression
factor dictated by the spontaneously broken flavor symmetry. As
already mentioned in the introduction, the large top mass is therefore
generically difficult to accommodate in this framework \cite{sss}.

\section{Generalization to arbitrary representations}
\label{sec:perglistringhisti}

In this section, we generalize the construction discussed above to arbitrary
representations of the electroweak and flavour groups. 

We generalize the minimal choice of ref.~\cite{sss} by
taking $\p^d$ and $\p^u$ to belong respectively to the 
${\bf (n_W^d\!+1)(n_W^d\!+2)/2}$ ($n_W^d$ times symmetric) and 
${\bf (n_W^u\!+1)(n_W^u\!+2)/2}$ ($n_W^u$ times symmetric) of $SU(3)_{W}$; 
the ${\bf 3}$ and ${\bf 6}$ that were used in ref.~\cite{sss} and
in the previous discussion correspond 
to the particular cases $n_W^d=1$ and $n_W^u=2$. Moreover, we take 
these fields to belong to the ${\bf 2j_F\!+1}$ (spin-$j_F$) 
representation of $SU(2)_F$, so that there are now $2j_F\!+1$ replicas of
them with identical $SU(2)_L \times U(1)_Y$ quantum numbers but different
$U(1)_F$ charges. Summarizing, we have bulk fields in the following 
representations of $SU(3)_W \times SU(2)_F$:
\begin{eqnarray}
\begin{array}{ccc}
\vspace{1em}
\displaystyle{\psi^l,\tilde \psi^l : 
\Big({\bf \frac {(n_W^l\!+1)(n_W^l\!+2)}2}, {\bf 2j_F\!+1}\Big)} \,,
\end{array}
\end{eqnarray}

The decomposition of the above general representations of the 
$SU(3)_W \times SU(2)_F$ group under its $SU(2)_L \times U(1)_Y \times U(1)_F$
subgroup, which we need to determine the coupling of the bulk fields 
to the brane fields, has the following form:
\begin{eqnarray}
\Big({\bf \frac {(n_W^l\!+1)(n_W^l\!+2)}2}, {\bf 2j_F\!+1}\Big) 
\rightarrow \hspace{10pt} \oplus \hspace{-15pt} {\raisebox{15pt}{}
\raisebox{-8pt}{}}_{\!j_W = 0}^{n_W^l/2} \hspace{13pt} 
\oplus \hspace{-12pt} {\raisebox{15pt}{}
\raisebox{-8pt}{}}_{\!\!\!\!\!\!\!\!\!\!m_{j_F} = -j_F}^{j_F} 
({\bf 2j_W\!+1})_{j_W-n_W^l/3,m_{j_F}} \,.
\label{eq:gendecomp}
\end{eqnarray}
We get therefore a set of representations of $SU(2)_L$ with half-integer spins $j_W$
ranging from $0$ to $n_W^l/2$, canonically normalized $U(1)_Y$ charge
equal to $j_W-n_W^l/3$ and $U(1)_F$ charges $m_{j_F}$ ranging from $-j_F$ 
to $j_F$. The only components that have the right quantum numbers to couple 
to the brane fermions are the $SU(2)_L$ doublets and singlets with $j_W=1/2$ 
and $j_W = 0$, which have $U(1)_Y$ charge\footnote{Notice that these have 
automatically the right hypercharge to couple to the standard left-handed 
doublets and right-handed singlets only in the special case $n_W^d=1$ and 
$n_W^u=2$ chosen in ref.~\cite{sss}. For more general values of $n_W^d \neq 1$ 
and $n_W^u \neq 2$, one needs to assign to the bulk fields a non-vanishing 
charge under the extra $U(1)^\prime$ factor that is needed to tune the weak 
mixing angle, which is equal to $(n_W^d-1)/3$ for $\psi^d,\tilde \psi^d$ and 
$(n_W^u-2)/3$ for $\psi^u,\tilde \psi^u$. Notice however that unless these charges 
are opposite to each other, that is if $n_W^u + n_W^d = 3$, two different fields 
are needed to give mass to the $u$ and the $d$ quarks, due to the restrictions 
set by the $U(1)'$-invariance of the coupling to the left-handed quarks.} equal 
to $1/2-n_W^l/3$ and $-n_W^l/3$, and $U(1)_F$ charges ranging from $-j_F$ to $j_F$. 

The action of the orbifold projection and the SS twist on the bulk fermion 
fields can be easily deduced by using some simple group-theoretical techniques. 
In the electroweak sector, the completely symmetric representations of $SU(3)_W$ 
we are considering contain states with values of the two Cartan generators 
$T_W^3$ and $2\, T_W^8/\sqrt{3}$ that fill an equilateral triangle in the 
corresponding plane. This triangle is oriented with its tip at the bottom and one 
of his sides at the top and horizontal. It can be sliced in essentially two 
different ways in a sum of lines, corresponding to decompositions with respect 
to nonequivalent but isomorphic maximal subgroups. Slicing the $SU(3)_W$ 
representation horizontally in rows, one obtains the decomposition with respect 
to the $SU(2)_L \times U(1)_Y$ preserved by the orbifold projection, with 
generators $T_W^{1,2,3}$ and $T_W^8/\sqrt{3}$.  It is then clear that the generator 
$T_W^8/\sqrt{3}$ appearing in the orbifold projection has a definite value for each 
$SU(2)_L \times U(1)_Y$ representation appearing in the decomposition 
(\ref{eq:gendecomp}). More precisely, it acts as $j_W - n_W^l/3$ on the 
component with $SU(2)_L$ spin $j_W$. In matrix form, where these components 
are ordered in block with a fixed $j_W$ ranging from $n_W^l/2$ to $0$ in decreasing
order and sub-entries corresponding to $m_{j_W}$ ranging from $-j_W$ to $j_W$ in 
increasing order\footnote{This ordering of the states differs from the one used 
for the particular example of section 3.}, the orbifold twist has therefore the following form:
\begin{eqnarray}
P^{\bf (n_W^l\!+1)(n_W^l\!+2)/2}_W =
\mathrm{diag}(\underbrace{(-1)^{n_W^l},\dots,(-1)^{n_W^l}}_{n^l_W+1\;\; {\rm \footnotesize times}};\dots;1,1,
1;-1 , -1;1)\,.
\end{eqnarray}
Slicing the $SU(3)_W$ representation diagonally, that is parallel to one of the 
two non-horizontal sides of the triangle, one obtains the decomposition with respect 
to a different $SU(2)' \times U(1)'$ subgroup associated to the Scherk-Schwarz 
twist, with generators $T_W^{6,7}$, $(- T_W^3 + \sqrt{3}\, T_W^8)/2$ and 
$(- T_W^3 - T_W^8/\sqrt{3})/2$. For each state of the original representation, the 
$SU(2)'$ spin $j'$ and its third component $m_{j'}$ are related to the the $SU(2)_L$ 
spin $j_W$ and its third component $m_{j_W}$ by the relations 
$j' = (n_W^l - j_W - m_{j_W})/2$ and $m_{j'} = (- n_W^l + 3 j_W -m_{j_W})/2$. 
This decomposition is useful to determine the action of the generator 
$T_W^6$ appearing in the Scherk-Schwarz twist. Indeed, one can rewrite 
$T_W^6 = (T_W^+ + T_W^-)/2$ in terms of the raising and lowering operators 
$T_W^\pm = T_W^6 \pm i T_W^7$ of the $SU(2)'$ subgroup. These leave $j'$ 
unchanged and raise/lower $m_{j'}$ by $1$ unit, or equivalently, they raise/lower 
$j_W$ by $1/2$ unit and lower/raise $m_{j_W}$ by $1/2$ unit. 
The generator $T_W^6$ acts in a non-diagonal way on the decomposition
(\ref{eq:gendecomp}), but its matrix elements can be easily determined using the 
standard SU(2) results. Its diagonal form is also easily derived, thanks to the fact that
any generator of an SU(2) group has the same diagonal form, due to the fact that there
is  only one Cartan generator. In our case, the diagonal form of $T_W^6$ must coincide 
in from with the generator $(-T_W^3 + \sqrt{3}\,T_W^8)/2$ representing the third component of 
the $SU(2)'$ spin. In terms of the quantum numbers defined by the decomposition 
(\ref{eq:gendecomp}), the latter acts as $(-n_W^l +3 j_W - m_{j_W})/2$ on the $m_{j_W}$-th 
element of the spin $j_W$ component. 
In the same matrix notation as above, this means
\begin{eqnarray}
t^{\bf (n_W^l\!+1)(n_W^l\!+2)/2}_W =
\mathrm{diag}(0,\frac 12,\dots,\frac {n_W^l}2; \dots ;
-\frac {n_W^l}2+\frac 12,-\frac {n_W^l}2+1;-\frac {n_W^l}2)\,.
\end{eqnarray}

In the flavour sector, the situation is similar but much simpler, since we start
with an $SU(2)_F$ group. The generator $T_F^3$ appearing in the orbifold 
projection is just the third component of the $SU(2)_F$ spin, and acts therefore
as $m_{j_F}$ on the $m_{j_F}$-th component of the decomposition (\ref{eq:gendecomp}).
One then finds that the projection matrix $P_F$ acts as 
$(-1)^{j_F-m_{j_F}}$ on the $m_{j_F}$-th component of the representation. In matrix 
notation, where these components are ordered with decreasing $m_{j_F}$ ranging from 
$j_F$ to $-j_F$\footnote{Again, this ordering differs from the canonical one used 
for the particular example of section 3.}, the orbifold twist has therefore the following form:
\begin{eqnarray}
P^{\bf 2j_F\!+1}_F  = \mathrm{diag}(1,-1,1,-1,\dots)\,.
\end{eqnarray}
The generator $T_F^1$ appearing in the Scherk-Schwarz twist can be written
more usefully as $T_F^1= (T_F^+ + T_F^-)/2$ in terms of the raising and lowering operators 
$T_W^\pm = T_F^1 \pm i T_W^2$ of the $SU(2)_F$ subgroup. This allows to 
compute in a simple way any of its matrix elements. Its diagonal form must 
coincide with that of the Cartan generator $T_F^3$, which acts as $m_{j_F}$ on the 
$m_{j_F}$-th component of the decomposition (\ref{eq:gendecomp}). The 
diagonal form of the twist is therefore given by
\begin{eqnarray}
t^{{\bf 2j_F\!+1}}_F  = \mathrm{diag}\Big(j_F,j_F-1,\dots,-j_F+1,-j_F\Big)\,.
\label{tnF}
\end{eqnarray}

We now describe the general situation that can be achieved in this more 
generic setting, in order to illustrate the basic features of the 
construction and its peculiarities compared to standard 4D 
flavour models. 

\subsection{Lagrangian}

The structure of the Lagrangian is the same as in the previous section. The
couplings of the family triplets 
of left- and right-handed brane fields $\phi = Q_{L}, u_{R}, d_{R}$ and 
their conjugates $\phi^c = Q_{R}^c, \mbox{-}u_{L}^c, d_{L}^c$ to the bulk
fields 
$\psi^l$ or $\tilde \psi^l$ are parametrized by family triplets of couplings
$e_1^{l}$ and $e_2^{l}$ with mass-dimension $1/2$, in each sector $l=u,d$. 
Each $\phi$ or $\phi^c$ can couple either to $\psi^l$ or $\tilde \psi^l$, and 
has therefore only one relevant coupling. To write these couplings more 
explicitly, it is convenient to embed the fields $\phi$ and $\phi^c$
into new fields $\Phi = Q, u, d, \tilde Q, \tilde u, \tilde d$ and their 
conjugates $\Phi^c = Q^c, u^c, d^c, \tilde Q^c, \tilde u^c, \tilde d^c$, which 
have the same matrix structure as the representations of $SU(3)_W \times
SU(2)_F$ to which the bulk fields they couple to belong, the extra entries being 
filled with zeroes\footnote{We denote the new embedded fields with the same 
letter as the original ones, but drop the $L,R$ subscripts to them.}.
The untilded and tilded fields in $\Phi$ or $\Phi^c$ contain 
those SM fermions $\phi$ or $\phi^c$ that have the right quantum numbers to 
couple to $\psi^l$ and $\tilde \psi^l$ respectively. With this notation, which 
is the appropriate generalization of the one used to deal with the particular 
example of sec.~3, the Lagrangian is obtained from eq.~(\ref{Lag}) by replacing 
the localized terms with
\begin{eqnarray}
\mathcal{L}^0 \!\!\!&=\!\!\!&
i \bar{Q} \gamma^{\mu} D_{\mu} Q 
+ i \bar{\tilde{Q}} \gamma^{\mu} D_{\mu} \tilde{Q} 
\nonumber \\ 
\!\!\!&\;\!\!\!& 
+\,\Big[\bar{Q}\,\hat e^{d}_{1}{}^\dagger \p^{d} 
+ \bar{\tilde{Q}}\,\hat e^{d}_{1}{}^\dagger \tilde{\p}^{d}
+ \bar{Q}^{c}\,\hat e^{u}_{1}{}^\dagger \p^{u} 
+ \bar{\tilde{Q}}{}^{c}\,\hat e^{u}_{1}{}^\dagger \tilde{\p}^{u} 
+ \rm{h.c.}\Big] \,, 
\label{locale1stringhe} \\
\mathcal{L}^{\pi R} \!\!\!&=\!\!\!&
i \bar{u}^{c} \gamma^{\mu} D_{\mu} u^{c} 
+ i \bar{\tilde{u}}{}^{c} \gamma^{\mu} D_{\mu} \tilde{u}^{c} 
+ i \bar{\tilde{d}} \gamma^{\mu} D_{\mu} \tilde{d} 
+ i \bar{d} \gamma^{\mu} D_{\mu} d 
\nonumber \\ 
\!\!\!&\;\!\!\!& 
+\, \Big[\bar{d}\, \hat e^{d}_{2}{}^\dagger\p^{d} 
+ \bar{\tilde{d}}\, \hat e^{d}_{2}{}^\dagger\tilde{\p}^{d}
+ \bar{u}^{c}\,\hat e^{u}_{2}{}^\dagger\p^{u} 
+ \bar{\tilde{u}}{}^{c}\,\hat e^{u}_{2}{}^\dagger \tilde{\p}^{u} 
+ \rm{h.c.}\Big] \,. 
\label{locale2stringhe}
\end{eqnarray}

To be more precise about the embeddings, let us denote $SU(2)_L \times U(1)_Y$ 
and family indices by $\alpha,\beta,\dots$ and $I,J,\dots=1,2,3$, and 
$SU(3)_W$ and $SU(2)_F$ indices by $i,j,\dots$ and $a,b,\dots$. The 
embedding of each field is then specified by some $(n_W^l\!+1)(n_W^l\!+2)/2$ 
by $2j_W\!+1$ matrix $({\cal I}_W)_{i\alpha}$ for gauge indices, where 
$j_W$ is $0$ for singlets and $1/2$ for doublets, and similarly by some 
$2j_F\!+1$ by $3$ matrix $({\cal I}_F)_{aI}$ for flavour indices. The position 
of each field $\phi$ or $\phi^c$ in $\Phi$ or $\Phi^c$ is uniquely determined 
by its $SU(2)_L\times U(1)_Y$ and $U(1)_F$ quantum numbers in the gauge and 
flavour sectors respectively. For the couplings, the embedding is trivial for 
gauge indices and is determined in an obvious way in terms of that of the
fields 
for flavour indices: it is a diagonal $2j_F\!+1$ by $2j_F\!+1$ matrix whose 
non-zero entries are the couplings that are relevant for each field, in the 
corresponding positions.

The embedding in the gauge sector generalizes the one used in ref.~\cite{sss}. 
Rather than reporting the matrices ${\cal I}_W$ for each field, we can exhibit 
the same information by reporting the expressions of the fields 
$\Phi_W = {\cal I}_W^\Phi \phi$ and $\Phi_W^c = {\cal I}_W^{\Phi^c} \phi^c$. 
These are $(n_W^l\!+1)(n_W^l\!+2)/2$-dimensional vectors will all 
the entries set to zero apart from the last three, which host the SM fields:
\begin{eqnarray}
\begin{array}{cccc}
Q_W = \tilde{Q}_W = \left(
\begin{array}{c}
0 \\
\vdots \\
0 \\
u_L \\
d_L \\
0 
\end{array}
\right)\,,\quad & 
d_W = \tilde{d}_W = \left(
\begin{array}{c}
0 \\
\vdots \\
0 \\
0 \\
0 \\
d_R 
\end{array}
\right)\,, 
\end{array}
\label{embedF1}
\end{eqnarray}
\vskip -15pt
\begin{eqnarray}
\begin{array}{cccc}
Q^{c}_W = \tilde{Q}^{c}_W = \left(
\begin{array}{c}
0 \\
\vdots \\
0 \\
\!d_{R}^{c}\! \\
\!\mbox{-}u_{R}^{c}\! \\
0 
\end{array}
\right)\,,\quad & 
u^{c}_W = \tilde u^{c}_W = \left(
\begin{array}{c}
0 \\
\vdots \\
0 \\
0 \\
0 \\
\!\mbox{-}u_{L}^{c}\!
\end{array}
\right) \,.
\end{array}
\label{embedF2}
\end{eqnarray}
The embedding in the flavour sector is done in a similar way and 
depends on the choice of flavour quantum numbers. Again, rather 
than reporting the matrices ${\cal I}_F$ for each field, one can 
consider directly the redefined fields $\Phi_F = {\cal I}_F^\Phi \phi$ 
and $\Phi_F^c = {\cal I}_F^{\Phi^c} \phi^c$. For $\Phi_F$, each 
SM fermion $\phi$ is embedded at the $(j_F-q_F+1)$-th entry if its flavour 
charge is $q_F$, and appears only in the untilded or tilded redefined 
fields if $j_F-q_F$ is respectively even or odd. Similarly, for the 
conjugate $\Phi_F^c$, each conjugate SM fermion $\phi^c$ is embedded at 
the $(j_F+q_F+1)$-th entry if its flavour charge is $-q_F$, and appears 
only in the untilded or tilded redefined fields if $j_F-q_F$ is 
respectively even or odd. As a consequence, for the embedding of the SM 
fields $\phi$ in $\Phi$, only the odd and even entries of respectively the 
untilded and the tilded redefined fields are relevant, all the other 
being always zero; for the embedding of the conjugate SM fields $\phi^c$ 
in $\Phi^c$, the situation is similar, and $\Phi^c$ is obtained from 
$\Phi$ through a reflection. Schematically, the structure is as follows,
with at most three non-vanishing entries for each vector:
\begin{eqnarray}
\begin{array}{cccc}
Q_F = \left(
\begin{array}{c}
0 \\
\vdots \\
0 \\
\!\!(Q_L)_{I_1}\!\! \\
0 \\
\vdots \\
0 \\
0
\end{array}
\right)\,, & 
\tilde{Q}_F = \left(
\begin{array}{c} 
0 \\
0 \\
\vdots \\
0 \\
\!\!(Q_L)_{\tilde I_1}\!\! \\
0 \\
\vdots \\
0
\end{array}
\right)\,, &
d_F = \left(
\begin{array}{c}
0 \\
\vdots \\
0 \\
\! (d_L)_{J_1} \! \\
0 \\
\vdots \\
0 \\
0
\end{array}
\right)\,, & 
\tilde{d}_F = \left(
\begin{array}{c}
0 \\
0 \\
\vdots \\
0 \\
\! (d_L)_{\tilde J_1} \! \\
0 \\
\vdots \\
0
\end{array}
\right)\,,
\end{array}
\label{embedu}
\end{eqnarray}
\vskip -10pt
\begin{eqnarray}
\begin{array}{cccc}
Q^{c}_F = \left(
\begin{array}{c}
0 \\
0 \\
\vdots \\
0 \\
\!\! (Q_R^c)_{I_1} \!\! \\
0 \\
\vdots \\
0
\end{array}
\right)\,, & 
\tilde{Q}^{c}_F = \left(
\begin{array}{c}
0 \\
\vdots \\
0 \\
\!\! (Q_R^c)_{\tilde I_1} \!\! \\
0 \\
\vdots \\
0 \\
0
\end{array}
\right) \,, & 
u^{c}_F =  \left(
\begin{array}{c}
0 \\
0 \\
\vdots \\
0 \\
\!\!\!(\mbox{-}u_L^c)_{K_1}\!\! \\
0 \\
\vdots \\
0
\end{array}
\right)\,, & 
\tilde{u}^{c}_F = \left(
\begin{array}{c}
0 \\
\vdots \\
0 \\
\!\!\!(\mbox{-}u_L^c)_{\tilde K_1}\!\! \\
0 \\
\vdots \\
0 \\
0
\end{array}
\right)\,.
\end{array}
\label{embedd}
\end{eqnarray}
In this expressions, $I_1$, $J_1$, $K_1$ and $\tilde I_1$, $\tilde J_1$,
$\tilde K_1$ are restricted family indices running respectively over those 
families for which the left-handed doublets, the right-handed down singlets
and the right-handed up singlets are embedded in untilded and tilded vectors.

Finally, the brane-bulk couplings are correspondingly embedded into diagonal 
matrices $\hat e^{l}_{1,2}$ at those entries that correspond to a 
non-vanishing entry of the redefined fields. They have
the following schematic form, with three non-vanishing entries labeled
by a family index $M$:
\begin{eqnarray}
\hat e^l_{1,2} = \mathrm{diag}(0,\dots,0,(e^{l}_{1,2})_{M_1},0,\dots,0,
(e^{l}_{1,2})_{M_2},0,\dots,0,(e^{l}_{1,2})_{M_3},0,\dots,0)\,.
\end{eqnarray}

\subsection{Fermion masses and mixings}
\label{sec:masseperstringhisti}

{}From the Lagrangian above one can proceed exactly as in
Sec.~\ref{sec:prototype} to derive the effective Lagrangian for the SM fermions,
which still has the form of eq.~(\ref{lagrph}). The general 
expressions of $\mathcal M$ and $\mathcal Z$ depend on the matrix elements of
the generator $T_W^6$ implementing the electroweak symmetry breaking and those
of arbitrary powers of the generator $T_F^1$ implementing the flavour symmetry 
breaking, which appear in the functions of eqs.~(\ref{fggapprossimate}).
The relevant matrix element of $T_W^6$ is universal and can be computed 
in general. It is the one connecting the next-to-last element of the 
embedding vector of the left-handed fields and their conjugates,
that is the $m_{j_W} = -1/2$ component of the doublet with $j_W = 1/2$,
and the last element of the embedding vector of the right-handed fields and
their conjugates, that is the singlet with $m_{j_W} = 0$ and $j_W = 0$. 
As already explained, this can be easily evaluated by 
rewriting $T_W^6 = (T_W^+ + T_W^-)/2$ in terms of the raising and lowering 
operators $T_W^\pm = T_W^6 \pm i T_W^7$  of the $SU(2)'$ subgroup defined
by the twist, which have non-vanishing matrix elements between neighbour 
states, namely $\sqrt{(j' \mp m_{j'})(j' \pm m_{j'} + 1)}$.
The matrix element we are interested in is therefore an ordinary
transition from the component with $m_{j'} = -n_W^l/2$ to the 
component with $m_{j'} = -n_W^l/2+1$ of an $SU(2)'$ representation of spin 
$j'= n_W^l/2$, and gives a factor $\sqrt{n_W\!\!\!\!{\raisebox{6pt}{}}^l}\;/2$.
The matrix elements of a generic power of $T_F^1$ can be computed similarly.
Here we simply rewrite the flavour traces in terms of the $2j_F\!+1$ by $3$
matrices 
${\cal I}_F$ defining how the family triplet of each SM field is embedded 
into an $(2j_F\!+1)$-dimensional flavour vector. The results are given by the 
following expressions:
\begin{eqnarray}  
\begin{array}{lll}
{\cal Z}_{L}^{d} \!\!\!&=\!\!\!& \displaystyle{
\mathbf{1} + \mathcal{E}_{1}^{d\dagger} \Big[
{\cal I}_F^{Q}{}^\dagger f(x_d, T^1_F \beta)\, {\cal I}_F^{Q} 
+ {\cal I}_F^{\tilde Q}{}^\dagger f(x_d, T^1_F \beta)\, {\cal I}_F^{\tilde Q} 
\Big] \mathcal{E}_{1}^{d}} \medskip\ \\
\!\!\!&\;\!\!\!& +\,\displaystyle{
\mathcal{E}_{1}^{u\dagger} \Big[
{\cal I}_F^{Q^c}{}^\dagger f(x_u, T^1_F \beta)\, {\cal I}_F^{Q^c} 
+ {\cal I}_F^{\tilde Q^c}{}^\dagger f(x_u, T^1_F \beta)\, {\cal I}_F^{\tilde
Q^c} 
\Big] \mathcal{E}_{1}^{u}} \,, \medskip \\ 
{\cal Z}_{L}^{u} \!\!\!&=\!\!\!& \displaystyle{
\mathbf{1} + \mathcal{E}_{1}^{u\dagger} \Big[
{\cal I}_F^{Q^c}{}^\dagger f(x_u, T^1_F \beta)\, {\cal I}_F^{Q^c} 
+ {\cal I}_F^{\tilde Q^c}{}^\dagger f(x_u, T^1_F \beta)\, {\cal I}_F^{\tilde
Q^c} 
\Big] \mathcal{E}_{1}^{u}} \medskip\ \\
\!\!\!&\;\!\!\!& +\, \displaystyle{\mathcal{E}_{1}^{d\dagger} \Big[
{\cal I}_F^{Q}{}^\dagger f(x_d, T^1_F \beta)\, {\cal I}_F^{Q} 
+ {\cal I}_F^{\tilde Q}{}^\dagger f(x_d, T^1_F \beta)\, {\cal I}_F^{\tilde Q} 
\Big] \mathcal{E}_{1}^{d}} \;,  \medskip \\ 
{\cal Z}_{R}^{d} \!\!\!&=\!\!\!& \displaystyle{
\mathbf{1} + \mathcal{E}_{2}^{d\dagger} \Big[
{\cal I}_F^{d}{}^\dagger f(x_d, T^1_F \beta)\, {\cal I}_F^{d} 
+ {\cal I}_F^{\tilde d}{}^\dagger f(x_d, T^1_F \beta)\, {\cal I}_F^{\tilde d} 
\Big] \mathcal{E}_{2}^{d}} \,, \medskip \\
{\cal Z}_{R}^{u} \!\!\!&=\!\!\!& \displaystyle{ 
\mathbf{1} + \mathcal{E}_{2}^{u\dagger} \Big[
{\cal I}_F^{u^c}{}^\dagger f(x_u, T^1_F \beta)\, {\cal I}_F^{u^c} 
+ {\cal I}_F^{\tilde u^c}{}^\dagger f(x_u, T^1_F \beta)\, {\cal I}_F^{\tilde
u^c} 
\Big] \mathcal{E}_{2}^{u}}\,, \raisebox{20pt}{} 
\end{array}
\label{Zgen}
\end{eqnarray}
and
\begin{eqnarray}
\begin{array}{lll}
\mathcal{M}^{d} \!\!\!&=\!\!\!& \displaystyle{
\sqrt{n_W^d}\,\mathcal{E}_{1}^{d}{}^\dagger \Big[
{\cal I}_F^{Q}{}^\dagger g_1(x_d, T^1_F \beta)\, {\cal I}_F^{d}
- {\cal I}_F^{\tilde Q}{}^\dagger g_1(x_d, T^1_F \beta)\, {\cal I}_F^{\tilde d}}
\medskip \\ \!\!\!&\;\!\!\!& \hspace{50pt} \displaystyle{
+\, {\cal I}_F^{Q}{}^\dagger g_2(x_d, T^1_F \beta)\, {\cal I}_F^{\tilde d} 
+ {\cal I}_F^{\tilde Q}{}^\dagger g_2(x_d, T^1_F \beta)\, {\cal I}_F^{d}  
\Big] \mathcal{E}_{2}^{d}\,m_W \,,} \medskip \\ 
\mathcal{M}^{u} \!\!\!&=\!\!\!& \displaystyle{
\sqrt{n_W^u}\,\mathcal{E}_{1}^{u}{}^\dagger \Big[
{\cal I}_F^{Q^c}{}^\dagger g_1(x_u, T^1_F \beta)\, {\cal I}_F^{u^c}
- {\cal I}_F^{\tilde Q^c}{}^\dagger g_1(x_u, T^1_F \beta)\, 
{\cal I}_F^{\tilde u^c}} 
\medskip \\ \!\!\!&\;\!\!\!& \hspace{50pt} \displaystyle{
+\,{\cal I}_F^{Q^c}{}^\dagger g_2(x_u, T^1_F \beta)\, {\cal I}_F^{\tilde u^c} 
+ {\cal I}_F^{\tilde Q^c}{}^\dagger g_2(x_u, T^1_F \beta)\, {\cal I}_F^{u^c} 
\Big] \mathcal{E}_{2}^{u}\,m_W \,.}
\end{array}
\end{eqnarray}

Once the above quantities have been computed, the physical implications of 
the Lagrangian (\ref{lagrph}) are uniquely determined and can be analyzed 
as follows. First, one performs a suitable redefinition of the fermions 
fields to reabsorb the non-trivial wave function factor and canonically 
normalize their kinetic terms. In this process, the mass matrices will 
however be changed to new matrices $\hat{\cal M}^l$. Second, one proceeds 
as in the SM and diagonalizes these two mass matrices through some unitary 
transformations ${\cal U}_{L,R}$ and ${\cal D}_{L,R}$ in the $u$ and $d$ 
sectors. This will then induce a CKM mixing matrix given by 
$V_{CKM} = {\cal U}_L^\dagger {\cal D}_L$. 

The above procedure is complicated by the
non-diagonal field redefinition that is required to get read of the 
wave function. One might fear that the new mass matrices $\hat{\cal M}^l$ 
that are generated after wave-function renormalization might have hierarchical
structures in powers of $\lambda$ that are messed up compared to those of ${\cal
M}^l$. However, as shown in general in Sec.~\ref{structure}, this is not the
case: at most the order one coefficients multiplying the powers of $\lambda$ in
the various entries are changed. We now generalize the discussion of
Sec.~\ref{sec:masses} for generic representations.

\subsubsection*{$\mathbf{x_l \gg 1}$}
\label{sec:xgg1perstringhisti}

In the limit of $x_l \gg 1$, the functions in eqs.~(\ref{fggapprossimatebis}) 
simplify to the form in eq.~(\ref{eq:xgg1}).
As we have seen, at leading order in $e^{-x_l}$ the wave functions reduce to 
diagonal constants:
\begin{eqnarray}  
{\cal Z}_{L}^{l} \simeq
\mathbf{1} + \frac{1}{x_d} \mathcal{E}_{L}^{d\dagger} \mathcal{E}_{L}^{d}
+ \frac{1}{x_u} \mathcal{E}_{L}^{u\dagger} \mathcal{E}_{L}^{u} \,,\qquad 
{\cal Z}_{R}^{l} \simeq 
\mathbf{1} + \frac{1}{x_l} \mathcal{E}_{R}^{l\dagger}\mathcal{E}_{R}^{l} \,.
\end{eqnarray}
The masses ${\cal M}^l$ take instead the form
\begin{eqnarray}
\mathcal{M}^{l} \simeq \displaystyle{\sqrt{n_W^l}\,e^{- x_{l}}
\mathcal{E}_{L}^{l}{}^\dagger \widetilde{Y}^{l} \mathcal{E}_{R}^{l}\,m_W} \,,
\end{eqnarray}
where $\widetilde{Y}^l$ are two $3 \times 3$ matrices that are functions of $\lambda$ 
and carry all the information about the group-theoretical details of the 
flavour sector. Assuming that $\lambda \ll 1$, they have the form 
(\ref{expyukawa}), but with completely fixed numerical coefficients, 
which can be easily computed using the standard realization of the 
$SU(2)$ algebra in terms of raising and lowering operators. 
Further assuming, for simplicity and without loss of generality, that 
the flavour charges $l_I$ of the right-handed fields are larger than 
the charges $q_I$ of the left-handed fields, and recalling that 
$\lambda = \pi \beta$, the result is, modulo a sign:
\begin{eqnarray}
\widetilde{Y}^l_{IJ}  \simeq \prod_{k=1}^{l_J - q_I} 
\sqrt{1 + \frac {j_F\!-l_J}k}\sqrt{1 + \frac {j_F\!+q_I}k}\,
\lambda^{l_J - q_I} \,.
\label{Xleading}
\end{eqnarray}
The first subleading corrections to these expressions can be easily evaluated using
again creation and annihilation operators. The relative effect represented by these
corrections is of order $\lambda^2$, and its precise expression, modulo a sign, is 
given by
\begin{eqnarray}
\frac {\Delta \widetilde{Y}^l_{IJ}}{\widetilde{Y}^l_{IJ}} \simeq 
\sum_{k=0}^{l_J - q_I + 1} \frac {(j_F + q_I + k)(j_F - q_I - k + 1)}{(l_J-q_I+1)(l_J-q_I+2)}\,
\lambda^2\,.
\label{Xsubleading}
\end{eqnarray}
From this expression it is clear that there is an obstruction against increasing too 
much the spin $j_F$ of the representation of the bulk mediators at fixed flavour 
charges for the brane fields. Indeed, doing so increases the relative impact 
of the subleading terms and puts a limit on how large the parameter $\lambda$
can be at fixed $j_F$, or viceversa how large $j_F$ can be at fixed $\lambda$, without 
spoiling the simple idea that the Yukawa texture is fixed by the leading terms with
powers of $\lambda$ fixed by the charges. Notice for instance that in the extreme 
limit in which $j_F$ is much larger than all of the charges, one finds that the leading 
term (\ref{Xleading}) goes like $j_F/(l_J-q_I)! \,\lambda^{l_J-q_I}$ if $l_J \neq q_I$ 
and $1$ if $l_J = q_I$, whereas the relative subleading correction (\ref{Xsubleading}) 
goes like $j_F^2/(l_J-q_I+1)\, \lambda^2$. Requiring that the latter be much smaller 
than $1$ then implies that $j_F \ll \sqrt{l_J-q_I+1}/\lambda$.
For $\lambda \sim 10^{-1}$ and reasonable charges, one must then take 
$j_F \ll 10$. For $j_F \sim 3-4$, as in the examples that we shall study below, the 
subleading corrections represent therefore a significant error of about $10 \%$.

The physical quark Yukawa couplings are obtained by redefining the quark 
fields to reabsorb the wave-function corrections ${\cal Z}_{L,R}^{l}$. 
The physical mass matrices are then found to be (see eq.~(\ref{physmass})):
\begin{eqnarray}  
m^l \simeq \sqrt{n_W^l}\, x_l e^{-x_l}  
\sin \alpha_L^l\, \widetilde{Y}^l \sin \alpha_R^l\, m_W \,.
\label{masse}
\end{eqnarray}

\subsubsection*{$\mathbf{x_l \ll 1}$}
\label{sec:xll1perstringhisti}

In the limit of $x_l \ll 1$, the masses in eq.~(\ref{massette}) generalize as
\begin{eqnarray}
\mathcal{M}^{l} \simeq \sqrt{n_W^l} \frac 1{x_{l}^2} 
\epsilon_{L}^l\!\!{}^\dagger \epsilon_{R}^{l}\,m_W\,.
\end{eqnarray}
The physical quark masses emerging after canonical normalization are then 
found to be
\begin{eqnarray}
m^l \simeq \sqrt{n_W^l} \sin \alpha_L^l \sin \alpha_R^l\, m_W\,.
\label{massebisperstringhisti}
\end{eqnarray}

\section{Model building}
\label{minimal}

In this section, we apply the general construction developed so far to build 
viable flavour models. We present two illustrative examples that emphasize 
some important phenomenological aspects.

\subsection{Mixing angles and mass ratios}
\label{sec:settemezzi}

The model presented in sec.~\ref{sec:prototype} produces the correct
structure of powers of $\lambda$ for Yukawa couplings, but suffers from large
group-theoretical coefficients that spoil the success of the chosen texture.
The simplest way to solve this problem is to assign charges in such a way as
no $\mathcal{O}(\lambda^0)$ term is present. Then, all entries will have comparable
numerical coefficients, and the power expansion will be consistent. We can start
for instance from
\begin{eqnarray}
Y^{d} \sim \left(
\begin{array}{ccc}
\lambda^{6}  & \lambda^{5} & \lambda^{4} \\ 
\lambda^{5}  & \lambda^{4} & \lambda^{3} \\ 
\lambda^{3}  & \lambda^{2} & \lambda
\end{array}
\right)\,, \qquad
Y^{u} \sim \left(
\begin{array}{ccc}
\lambda^{7}  & \lambda^{6} & \lambda^{4} \\ 
\lambda^{6}  & \lambda^{5} & \lambda^{3} \\ 
\lambda^{4}  & \lambda^{3} & \lambda
\end{array}
\right)\,.
\label{parametrizzayukawa8}
\end{eqnarray}
The simplest flavour charge assignment for the brane fermions that is 
compatible with these textures is given by
\begin{eqnarray}
q_I = \Big\{\!-\!\frac 72,-\frac 52,-\frac 12\Big\} \,,\qquad
d_I = \Big\{\frac 52,\frac 32,\frac 12\Big\} \,,\qquad
u_I = \Big\{\frac 72,\frac 52,\frac 12\Big\} \,.
\end{eqnarray}
Since the maximal absolute value of the charge is now $7/2$, the smallest 
allowed representation for the bulk fermions has now spin $j_F=7/2$.

Assuming as before $x_l \gg 1$ to simplify the analysis of the effects
on order one coefficients due to wave-function corrections, the induced 
mass matrices $\mathcal{M}^u$ and $\mathcal{M}^d$ are given by 
eqs.~(\ref{mamd}) and (\ref{mamu})
with:
\begin{eqnarray}  
 \widetilde{Y}^{d}  \!\!\!&=\!\!\!& 4\,\lambda \times  \left( 
\begin{array}{ccc}
\frac{\sqrt{7}}{4}{\lambda}^5 & 
-\frac{\sqrt{21}}{4}{\lambda}^4 & 
-\frac{\sqrt{35}}{4}{\lambda}^3 \\[1mm]
-\frac{3}{2}{\lambda}^4 & 
\frac{5 \,{\sqrt{3}}}{4}{\lambda}^3 & 
{\sqrt{5}}\,{\lambda}^2 \\[1mm]
{\sqrt{5}}\,{\lambda}^2 & 
- \frac{\sqrt{15}}{2}\lambda & 
-1
\end{array}
\right) \;, \label{babdotto} \\
\widetilde{Y}^{u}  \!\!\!&=\!\!\!&  4 \, \lambda \times \left(
\begin{array}{ccc}
\frac{1}{4} \lambda^6 &   
\frac{\sqrt{7}}{4}\,\lambda^5  & 
-\frac{\sqrt{35}}{4}\,\lambda^3  \\[1mm]
-\frac{\sqrt{7}}{4}\,\lambda^5 &
-\frac{3}{2}\lambda^4  & {\sqrt{5}}\,\lambda^2 \\[1mm]
\frac{\sqrt{35}}{4}\,\lambda^3   &
{\sqrt{5}}\,\lambda^2  & -1 
\end{array}
\right) \;,
\label{babuotto}
\end{eqnarray}
The mass matrices that are obtained in this case still have the problem
of a too low overall scale, but it is now possible to reproduce mass ratios 
and mixing angles with reasonable values of the parameters (except for the 
down quark mass which is too low).

\subsection{Example with improved overall scale}
\label{sec52}

The problem of the small overall scale can be solved by introducing, 
in addition to a pair of bulk fermions that are flavour-charged and induce 
general hierarchical mass matrices, an extra pair of bulk fermions that 
are flavour-neutral and contribute therefore only to the mass of 
flavour-neutral states. Assigning third-generation quarks a vanishing charge, 
neutral bulk fermions will only contribute to the $(3,3)$ entries of quark
masses. If charged bulk fermions are heavier, all the other entries will be
additionally suppressed by a factor $e^{-\pi R (M_l^C-M_l^N)}$, where $M_l^C$ and
$M_l^N$ stand for the masses of charged and neutral bulk fermions respectively.
It is clear that in this case the mass ratio between the third and the first
two generations is not a prediction of the flavour model any more, but stems 
from the exponential factor $e^{-\pi R (M_l^C-M_l^N)}$. Taking into account 
this extra suppression, we can choose for example
\begin{eqnarray}
Y^{d} \sim \left(
\begin{array}{ccc}
\lambda^{5}  & \lambda^{4} & \lambda^{3} \\
\lambda^{4}  & \lambda^{3} & \lambda^2 \\ 
\lambda^{2}  & \lambda & 1
\end{array}
\right)\,, \qquad
Y^{u} \sim \left(
\begin{array}{ccc}
\lambda^{6}  & \lambda^{4} & \lambda^{3} \\
\lambda^{5}  & \lambda^{3} & \lambda^2 \\
\lambda^{3}  & \lambda & 1
\end{array}
\right)\,.
\label{parametrizzayukawa7}
\end{eqnarray}
The simplest flavour charge assignment for the brane fermions that 
realize these is 
\begin{eqnarray}
q_I = \big\{\!-\!3,-2,0\big\} \,,\qquad
d_I = \big\{2,1,0\big\} \,,\qquad
u_I = \big\{3,1,0\big\} \,.
\end{eqnarray}
The smallest allowed representation for the charged bulk fermions has 
in this case spin $j_F=3$. 

These charged states give a contribution to the mass matrices 
$\mathcal{M}^u$ and $\mathcal{M}^d$ given by 
eqs.~(\ref{mamu}) and (\ref{mamd}) with 
\begin{eqnarray}
\widetilde{Y}^{d} \!\!\!&\simeq\!\!\!& \left( 
\begin{array}{ccc}
-\sqrt{6}\,{\lambda}^5 & 
\sqrt{15}\,{\lambda}^4 & 
2\sqrt{5}\,\,{\lambda}^3 \\[1mm]
-5\,{\lambda}^4 & 
2\sqrt{10}\,{\lambda}^3 & 
\sqrt{30}\,{\lambda}^2 \\[1mm]
-2\sqrt{3}\,{\lambda} & 
\sqrt{30}\,\lambda & 
-1
\end{array}
\right) \,,\label{babd7} \\
\widetilde{Y}^u \!\!\!&=\!\!\!& \left(
\begin{array}{ccc}
-\lambda^6 &  
\sqrt{15}\,\lambda^4 & 
2\sqrt{5}\,\lambda^3 \\[1mm]
-\sqrt{6}\,\lambda^5 &
2\sqrt{10}\,\lambda^3 & 
\sqrt{30}\,\lambda^2 \\[1mm]
2\sqrt{5}\,\lambda^3 &
-2\sqrt{3}\,\lambda & 
-1
\end{array}
\right)
\;.
\label{babu7}
\end{eqnarray}
For the corresponding flavour-neutral states, if we stick to the $SU(3)_W$
representations used in ref.~\cite{sss}, we still have a problem with the top
mass, which remains too low. As an illustrative example, one can choose a
rank $6$ symmetric representation for the flavour-neutral fermion coupling to
the top quark, even though one should check that the cutoff is not lowered too
much by the presence of fermions in large representations of $SU(3)_W$. 
With this caveat, the situation improves, and we can reproduce all masses 
and mixing angles with reasonable values of the parameters, except again 
for the down quark which tends to be too light.

\subsection{FCNC processes and CP violation}
\label{CPV}

Since there is a mixing between brane and bulk fermions, tree-level FCNC
couplings to the $Z$ boson are expected to arise. On general grounds, they will
be suppressed by $\alpha^2$ and by the appropriate power of $\beta$. It remains
to be seen whether in any specific model this suppression is sufficient to
guarantee a successful description of FCNC phenomena: to this aim, we are
presently carrying out a full one-loop phenomenological analysis.

In all the above discussions, for simplicity, we have taken the $\epsilon$
couplings to be real. In general, they are complex numbers and their phases
enter the effective mass matrices and the CKM matrix. The strength of CP
violation then depends on the size and phases of $\epsilon$ parameters, and can
be estimated in any specific model.
 
\section{Conclusion}
\label{conclusion}

We have proposed a mechanism to implement flavour symmetries in gauge-Higgs
unification models. In five-dimensional orbifold constructions the only possibility
consists in a flavour $SU(2)_F$ symmetry broken to $U(1)_F$ by the orbifold
projection and then to nothing via a compactification twist. Assuming that the
problems connected to electroweak symmetry breaking in gauge-Higgs unification
were solved, our proposal can successfully predict the orders of magnitude of all
mass ratios and mixing angles. Quantitative agreement can be obtained with
reasonable values of all relevant parameters. We stress that this class of
models is much more constrained than ordinary FN abelian flavour models because
of the higher-dimensional non-Abelian nature of the flavour symmetry. We are
presently investigating the phenomenology of FCNC processes in this kind of
construction, both at the tree and the one-loop levels. 

An interesting possibility would be to implement our idea in the framework of
warped five-dimensional models or in six-dimensional orbifolds, in which 
electroweak symmetry breaking seems more successful (see for instance 
\cite{contino,hosotani} and \cite{sssw,gaugehiggsorb2}).

\section*{Acknowledgments}

We would like to thank S.~Lavignac, R.~Rattazzi, A.~Romanino, M.~Serone and 
A.~Wulzer for useful discussions. We also thank the Theory Division of CERN 
for hospitality. This work has been partly supported by the European 
Commission through a Marie Curie fellowship, under the contract 
MRTN-CT-2004-005104 and through the programme "The quest for 
unification" under contract MRTN-CT-2004-503369, 
the Swiss National Science Foundation, the German 
Bundesministerium f\"ur Bildung und Forschung under the contract 05HT4WOA/3, 
and the German-Israeli Foundation under the contract G-698-22.7/2002.
M.S. acknowledges MECD for financial support through FPU fellowship
AP2003-1540.

\end{document}